\documentclass[aps,showpacs,floats,
superscriptaddress,nofootinbib]{revtex4}  
\newcommand{\gsim}{ \mathop{}_{\textstyle \sim}^{\textstyle >} }
\newcommand{\lsim}{ \mathop{}_{\textstyle \sim}^{\textstyle <} }
\newcommand{\ETmiss}{ {\not\!\!E}_{\rm T} }
\usepackage{graphicx}
\begin{document}
\large

\preprint{}
\preprint{KEK-TH 1088}
\title{Discovery of supersymmetry with degenerated mass spectrum}
\author{Kiyotomo Kawagoe }
\affiliation{Department of Physics, Kobe University, Kobe,657-8501, Japan }
\author{Mihoko M. Nojiri }
\affiliation{IPNS, KEK, Oho 1-1  305-0801, Japan}
\date{\today}
\begin{abstract}
Discovery of supersymmetric (SUSY) particles 
at the Large Hadron Collider (LHC) has been studied 
for the models where squarks and gluino are much heavier 
than the lightest supersymetric particle (LSP). 
In this paper, we investigate the SUSY discovery in the models with 
degenerated mass spectrum up to $m_{\rm LSP}\lsim 0.7 m_{\tilde{q}}$. 
Such mass spectrum is predicted in certain parameter region of the mixed 
modulas anomaly mediation (MMAM) model. 
We find that the effective transverse mass of the 
signal for the degenerated parameters shows the distribution 
similar to that of  
the background. Experimental sensitivity  of the SUSY particles at the  LHC 
therefore depends on the uncertainty of the background in this class of model. 
We also find that SUSY signal shows an interesting correlation between 
$M_{\rm eff}$ and $\ETmiss$ which may be used to 
determine the signal region properly to enhance the S/N ratio even if 
the sparticle masses are rather degenerated.  The structure 
is universal for the models with  new heavy colored particles 
decaying  into visible particles 
and a stable neutral particle, dark matter. 
\end{abstract}

\maketitle 
\section{Introduction: Transverse Physics and its 
non-transverse limit}

Minimal supersymmetric standard model (MSSM)
is one of the most promising candidates
of the physics beyond the standard model (SM) that
may  solve the
hierarchy problem in the Higgs sector~\cite{MSSM}.
The model predicts a set of superpartners (sparticles)
whose charges are exactly the same to that of the SM partners
but the spins are different by one half.

These sparticles will be directly searched at
the Large Hadron Collider (LHC) at CERN, which is
a $pp$ collider with 14~TeV
center of mass energy.  
The LHC is scheduled to
start in 2007.  One of the interesting
features of the MSSM  model is the conserved  R parity, and
all sparticles are assigned to have odd R parity. Because
of the conservation law, the
lightest supersymmetric particle (LSP)  is stable, and
is assumed as a good candidate for the dark matter.  
At collider experiments, sparticles
are produced in pairs, 
and a sparticle decay should produce
at least one  LSP.
SUSY events at collider experiments
therefore 
have a distinctive missing transverse momentum.

If  supersymmetry (SUSY) is an exact symmetry, a particle and its
superpartner have a same mass. Therefore  SUSY
must be broken spontaneously. The SUSY
is considered to be broken in
``hidden sector'', and the SUSY breaking will be mediated to
our sector by mediation fields. The mechanism is
already under severe constraints. For example, the
mass of the sfermions with same charge must
be universal so that they do not
cause dangerous FCNC, or sfermion masses
must be so heavy that correction from
the SUSY sector is suppressed.  This suggests
that yet unknown symmetry/dynamics
exists in the sector. The origin of the
supersymmetry
breaking and mediation mechanism may be understood
indirectly by measuring  masses of the SUSY particles.
Therefore both the discovery and mass determination of the
SUSY particle have been studied intensively in recent
years~\cite{TDR,HP,LHCLC}. 

If SUSY is broken at some high energy 
scale  and the boundary conditions are universal at the scale,
strongly interacting (SI) SUSY particles 
are  heavier than electroweakly interacting (EWI) SUSY particles. 
For example, 
gaugino masses 
follow the relation 
\begin{equation}
M_1: M_2: M_3\sim 0.4: 0.8: 2.4 
\end{equation}
in the minimal supergravity model, therefore the gluino is 
six times heavier than  the bino-like particles.  At the LHC, 
SI SUSY particles will be copiously produced 
in $pp$ collisions, and subsequently 
decay into lighter EWI SUSY particles. 

The gluino  and squark decays are  associated with 
jets with high transverse momentum ($p_T$).  
The transverse momentum is  the order of 
the gluino and squark masses. Moreover, because the LSP
 is significantly  lighter than the gluino,  the  
LSP from the gluino decay  also has a
high  $p_T$.  They would give a large missing transverse 
momentum to the SUSY events. In addition,  decays of the EWI sparticles may 
produce high $p_T$ leptons. 
Events from the standard model (SM) processes 
do not have such high $p_T$ particles. 

Motivated by these observations, 
following  cuts are often applied to reduce the 
SM background events to the SUSY signal events~\cite{TDR};
\begin{itemize}
\item An event is required to have at least one jet with $p_T>100$~GeV 
and three jets with $p_T>50$~GeV within  $\vert \eta\vert <3$, 
\item  The effective mass of the event must satisfy 
$M_{\rm eff}>400$~GeV, where the effective mass is defined 
using the transverse missing energy ($\ETmiss$) and the transverse momentum of
four leading jets as:
\begin{equation}
M_{\rm eff}\equiv\sum_{i=1,...4}p_{Ti}+ \ETmiss . 
\end{equation}
If the event has hard isolated  leptons, 
the effective mass may be defined as 
follows: 
\begin{equation}
M_{\rm eff}\equiv \sum_{i=1,...4} 
p_{Ti}+\sum_{\rm leptons} p_{Tl} + \ETmiss ,
\end{equation}
where sum of the lepton $p_T$ can be taken over the 
leptons with $p_T>20$~GeV and $\vert\eta\vert <2.5$~GeV. 
\item The missing transverse energy
must satisfy the relation:
\begin{equation}
\ETmiss >{\rm max}(0.2 M_{\rm eff}, 100{\rm GeV}).
\end{equation}
\item 
The transverse sphericity $S_T$ must be
greater than 0.2, where $S_T$ is defined as 
$2\lambda_2/(\lambda_1+\lambda_2)$, with $\lambda_1$ and 
$\lambda_2$ being the eigenvalues of the $2\times 2$ 
sphericity tensor $S_{ij} =p_{ki}p_{kj}$ formed by summing over the 
transverse momentum of all calorimeter cells. 
\end{itemize}
To reduce the background further, hard, isolated  lepton(s) 
may be required.  
These cuts are good enough to reduce the SM backgrounds from  
$t\bar{t}$+$n$-jets and $W(Z)$+$n$-jets productions 
down to a manageable 
level, although the production cross section of the SM 
processes may be  ${\cal O}(10^4)$ higher than signal cross sections.  
While the SUSY production section reduces very quickly 
as sparticle masses increase beyond 1~TeV, but the signature 
peaks at higher $M_{\rm eff}$  where backgrounds can be ignored . 
Previous studies show that the 
squark and gluino  with mass around 2.5~TeV 
can be found at the LHC in the minimal super gravity model (MSUGRA). 

In MSUGRA, the SM background 
after the cuts can be neglected safely.  Then, 
the distribution of accepted events are also useful 
to determine the mass scale of SUSY particles.  
For example, 
the peak of  the $M_{\rm eff}$ distribution is sensitive to the 
squark and gluino masses. 
For the events with same flavor opposite 
sign dileptons,  the invariant mass distributions,
$m_{ll}$, $m_{jl}$, and 
$m_{jll}$,
are useful to reconstruct the SUSY particle masses 
$m_{\tilde{\chi}^0_1}$, 
$m_{\tilde{\chi}^0_2}$, $m_{\tilde{q}^0_1}$
and $m_{\tilde{l}^0_1}$. 

Recently 
it is pointed out that  a string inspired model based 
on the flux compactification (KKLT models)~\cite{LINDE}  
predicts a mass relation different 
from that of the MSUGRA~\cite{CHOI, FLM,YAMA}. 
This is called mixed modulus anomaly mediation (MMAM)  model. 
This model 
has  a volume modulas $T$ and  a compensator 
field of minimum supergravity model $C$ 
as a messenger of the SUSY breaking. 
The SUSY mass spectrum depends on the ratio of 
the two SUSY breaking parameters 
$F_T$ and $F_C$. The unification 
scale of the sparticle masses depends on the ratio. 
It is interesting that the unification scale of the 
soft SUSY parameters can be much lower than 
the GUT scale in this model.  They may unify even at the weak scale 
for a special choice of the model parameters. 

When  sparticle masses are  degenerated  at the weak scale, 
we expect  a reduced  probability to have high $p_T$ jets,  
and  smaller $M_{\rm eff}$ and $\ETmiss$ for given 
squark and gluino masses.   
This means that the standard SUSY cuts reduce the signal 
events as well, and SUSY discovery is more 
affected by the SM background. 

Quantitative 
understanding of the SM background may be 
required in this case.  Existing background estimates 
have large theoretical uncertainty coming 
from the scale of the strong coupling. Recently, several 
groups has emphasized  importance to include 
the matrix element correction~\cite{MANGANO, RAINWATER, TEV4LHC} 
in the previous parton shower estimate of the background, which 
significantly change the background distribution in the signal region.
When overlap between the signal and background 
distribution is large,  as it is expected in the degenerated SUSY spectrum, 
the uncertainty of the background must be 
taken seriously, and we also  need to reconsider the cuts to 
reduce the background.

The purpose of this paper is to illustrate the phenomenology of the 
degenerated supersymmetry. We take the MMAM model as an example.
 By changing the ratio $F_C/F_T$,  
the mass spectrum changes smoothly from MSUGRA-like  one
to anomaly-mediation-like one. In between, there are 
regions of parameters where  squark, 
slepton, and gaugino masses are significantly degenerated 
compared with those expected in MSUGRA. 
The model therefore provides one dimensional parametrization 
from  the ``transverse signature''  to its non-transverse limit. 
The investigation of  our analysis may easily  be extended to 
other SUSY/non-SUSY scenarios with 
$\ETmiss$ signature,  such as the universal extra 
dimension model or little Higgs model with $T$ parity.

This paper is organized as follows. 
In section 2 we describe the 
MMAM model and its mass spectrum 
with an emphasis on the region of 
parameter space where sparticle masses are degenerated. 
In section 3, we describe our Monte Carlo simulations. 
In section 4, we study how SUSY event distribution 
depends on sparticle mass degeneracy and 
compare it with the background. We find
that the signal $M_{\rm eff}$ distribution 
is quite similar to that of the background if 
$m_{\tilde{\chi}^0_1}\gsim 0.5 m_{\tilde{q}, \tilde{g}}$. However 
we find a certain relation among $\ETmiss$  and $M_{\rm eff}$ 
 that is universal for the model 
with heavy colored particle that decays into the stable 
neutral particle 
 LHC will miss the SUSY signature if 
$m_{\tilde{\chi}^0_1}\gsim 0.7m_{\tilde{q}, \tilde{g}}$. 
Section 5 is devoted to discussions and conclusions. 

\section{The MMAM  model}\label{sec:model}

\subsection{Boundary conditions at GUT scale}

In this section, we briefly describe the MMAM 
model following the notation in \cite{CHOI} .
In this model, all shape modulus and dilaton will be fixed by non-zero 
flux  on a CY manifold in the Type IIB string theory. 
The low energy $N=1$ Lagrangian of this model is given by 
unfixed volume modulus $T$, compensator field $C$, and 
gauge and matter fields $W^a$ and $Q_i$ as 
\begin{eqnarray}
S_{N=1}&=&\int d^4x \sqrt{g^C}\left[ d^4\theta C C^*
(-3 \exp(-K_{\rm eff}/3)) \right. 
\cr
&& \left.
+\left\{
\int d^2\theta \left(f_a W^{a\alpha} W^a_{\alpha}
+ C^3 W_{\rm eff }\right)+ h.c.
\right\} \right]. 
\end{eqnarray}

Here $K_{\rm eff}= -3 \ln (T+T^*) + Z_i (T+ T^*) Q_i^* Q_i$, 
$g^c_{\mu\nu}$ is the 
4D metric in the super conformal frame. 
$T=T_0+F_T\theta^2$ and  $C= C_0+ F_C\theta^2 $ 
are the volume modulas and 
the chiral compensator superfield of $ N=1$ SUGRA,  respectively. 
$W_{\rm eff}=W_0(T) + \frac{1}{6}\lambda_{ijk} Q_iQ_jQ_k$ 
is the superpotential for $T$ and matter.  
$T$ dependent function $Z_i$ and  $f_i$ may be expressed as 
\begin{equation}
Z_i=\frac{1}{(T+T^{*})^{n_i}},\ \  f_a=T^{l_a},
\end{equation} 
where $n_i$  is the modular weights and   $n_i=0(1)$ 
for matter fields located on D7 (D3) branes, and 
$n=1/2$ for matter living at brane intersections~\cite{CHOI}. 

In KKLT model,  $W_0= w_0 -A \exp^{-aT}$, where 
the last term of $W_0$ expresses the non-perturbative 
effect such as the gaugino condensation in D7 brane, 
which fix the volume  modulus,  $w_0$ is the contribution 
of the flux. In addition to the $N=1$ supersymmetric 
action, there are  contributions from anti-D3 branes which break 
supersymmetry and uplift the potential from AdS vacuum to 
(nearly Minkowski) de Sitter vacuum.  The term is expressed by a 
spurion operator depending on $T$ and $C$,  
and minimum of the potential will be 
obtained by solving the effective $N=1$ action and the 
lifting potential. 

The resulting theory is parametrized by $F_C/C_0
\sim m_{3/2}$ and $F_T/(T+T^*)$  .
The SUSY breaking terms are obtained by expanding the action by $F_T$ and $F_C$. Here we define the soft terms as 
\begin{equation}
 L_{soft}=-\frac{1}{2} M_a\lambda^a\lambda^a 
 -m_i^2\vert \tilde{Q}_i\vert ^2 -A_{ijk}y_{ijk}
 \tilde{Q}_i\tilde{Q}_j\tilde{Q}_k +hc 
 \end{equation}
where $y_{ijk}$ is a canonically normalized Yukawa coupling
\begin{equation}
L_{\rm soft}=\frac{\lambda_{ijk}}{\sqrt{e^{-K_0}Z_iZ_jZ_k}}. 
\end{equation}
They are explicitly written as functions of $m_{3/2}$ and 
$R\equiv m_{3/2}(T+T^*)/F_{T}$ as follows;
\begin{eqnarray}
M_a&=& \left(\frac{l_a}{R}+ \frac{b_ag^2_{GUT}}
{16\pi^2}\right)m_{3/2}
\cr
m_i^2&=&\left(\frac{m_i}{R^2} + \frac{1}{R }\frac{\partial \gamma_i}{\partial \ln
T}
-\frac{1}{4}\frac{\partial\gamma_i}{\partial \ln \mu }\right) m_{3/2}^2
\cr
A_{ijk}&=&\left(\frac{1}{R}(m_i+m_j+m_k) -\frac{1}{2}(\gamma_i+\gamma_j+\gamma_k)\right)m_{3/2}
\label{eq:soft}
\end{eqnarray}
where $m_i=1-n_i$
\footnote{
Sign convention for the $A$ parameter 
 is such that the off-diagonal 
element of $\tilde{\tau}$ mass matrix  
is $-m_{\tau}(A_{\tau} +\mu \tan\beta)$. }
and
\begin{equation}
\gamma_r=\mu \frac{d\ln Z_r}{d\mu}= 
\frac{1}{8\pi^2} \left(2 \sum_aC^a_r g_a^2 -d_r y^2\right)
\end{equation}
with
$C_r=\sum_a T_a^2(r)$, namely 
for  matter in the fundamental representation
$C^3_F= 4/3$, $ C^2_F= 3/4$, 
$C'=Y^2$. 
We only include the effect of the top Yukawa coupling 
$y$,  therefore $d_{Q_3}= 1$, $d_T=2$, $d_{H2}=3$
and $d_i=0$ otherwise.

The scale dependence of $\gamma_i$ is expressed as 
\begin{equation}
\frac{d\gamma_i}{ d\mu}=\frac{b_a C^a_i}{32\pi^4}g_a^4 -d_i \frac{y^2}{32\pi^4}
\left(\frac{D}{2}y^2 -\sum_a C^a g^2_a\right) 
\end{equation}
where
\begin{eqnarray}
\mu \frac{dy^2}{d\mu} &=& \left(\frac{Dy^2}{2} -\sum_a C^a g_a^2\right) 
\frac{y^2}{4\pi^2}, 
\end{eqnarray} 
and 
\begin{eqnarray}
\frac{dg}{d\ln \mu}
=\frac{b_a}{8\pi^2} g^3. 
\end{eqnarray}
Here,  $D=\sum_r d_r=6$, $C^3=\sum_r C^3_r=\frac{8}{3}$, 
$ C^2=\frac{3}{2}$, $C'=\frac{13}{18}$, 
and $b'= 11$, $b_1= 33/5$ , $b_2=1$, $b_3= -3$. 

Finally, 
\begin{equation}
\frac{d\gamma_i}{d \ln T} = (m_i+m_j+ m_k) \frac{d_i}{8\pi^2} y^2
-\sum_a\frac{C^a_r}{4\pi^2}g_a^2.  
\end{equation}

\subsection{Mass spectrum in the MMAM  model} 

In Eq.~(\ref{eq:soft}), the highest term in $1/R$ is the 
contribution of modulas $T$, while the 
terms independent from $1/R$ is that of pure anomaly 
mediation. 
In \cite{CHOI}, it is pointed out that the mass spectrum in the 
MMAM  model shows a special 
feature if  $\alpha\sim 1 $, where 
\begin{equation}
\alpha=\frac{R}{\ln(M_{\rm pl}/m_{3/2})}.  
\end{equation}
This can be seen  by investigating  the low energy 
mass parameters as functions of $\alpha$. 
For example, 
gaugino masses  at the scale $\mu$  may be expressed as 
\begin{equation} 
M_a(\mu)=\frac{m_{3/2}}{R}
\left[
1-\frac{1}{4\pi^2} b_a g_a^2(\mu)\ln\left(
\frac{M_{GUT}}{(M_{Pl}/m_{3/2})^{\alpha/2} \mu}
\right)
\right]
\end{equation}
in one loop level. 
When $\alpha\sim 2$, the $\log$ term in the equation
becomes  0  at  $M_{SUSY}$. Namely, if $l_a$ does not 
depend on the gauge group, 
the  gaugino masses 
unify at the weak scale,  rather than at the GUT scale. 

The mass spectrum in  the matter sector 
depends on  $n_i$. If Yukawa couplings 
$y_{ijk}$ are non-vanishing only for the combination 
$Q_iQ_jQ_k$ satisfying  $n_i+n_j+n_k=2$, or if the 
effect of the Yukawa couplings in the RGE can be ignored,  
the scalar masses and trilinear couplings 
also unify at the same scale of the gaugino mass unification. 
This relation is satisfied for the choice 
 $n_H=1$  and $n_{matter}=1/2$ and  we call this choice 
 of the boundary condition as model A. 
Even 
if this condition is not satisfied, squark and slepton 
masses tend to be close with  each other  at $M_{SUSY} $ when 
gaugino masses unify at low energy scale, because 
the gaugino loop corrections to the sfermion masses 
are roughly equal.

In Table \ref{table:alphar}, 
we show the relation between $R$ and $\alpha$ 
for a fixed gluino mass at the GUT scale of 450~GeV. This 
corresponds to 
$M_{\tilde{g}}(\mu)=\alpha_s(\mu)/\alpha_{\rm (GUT)}\times
M_3{\rm (GUT)}=1.1$~TeV.  
We can see that the unification 
of gaugino or squark and slepton masses occurs at 
$R\sim 60$.  

\begin{table}[th]
\begin{center}
\begin{tabular}{|ccc|}
\hline
$\alpha$ & $m_{3/2}/R$(TeV) & $R$ \cr
\hline
  $0.26\times 10^{-2}$ & 0.45 &  0.1\cr
  0.30 & 0.50& 10\cr
  0.61 & 0.56 & 20\cr
  0.92 & 0.63&  30\cr
  1.25 & 0.73 & 40\cr
  1.58 & 0.86&  50\cr
  1.92 & 1.06&  60\cr
\hline
\end{tabular}
\caption{The relation between $\alpha$, $R$ and $m_{3/2}/R$(TeV) 
for $M_3{\rm (GUT)}=450$~GeV }
\label{table:alphar}
\end{center}
\end{table}

For  the following discussion, we calculate  the low energy mass spectrum 
using ISAJET~\cite{ISAJET} version 7.72
which solves the boundary 
condition given in section 2.1.  
The sparticle mass spectrum for the model A  is listed in 
Table \ref{table:massA}. 
As $\alpha$ increases, 
gaugino masses, and slepton and squark masses get 
closer, and the model shows the mass pattern 
different from  that of MSUGRA. 
The numerical values  roughly agree with  those in \cite{CHOI} 
\footnote{Note 
that ISAJET runs two loop RGE for all soft parameters, while 
the boundary conditions are calculated by the formula using one 
loop RGE.}.

\begin{table}[thb]
\begin{center}
\begin{tabular}{|c|ccccccc||cc|}
\hline
$R$ & $M_3$ &  $M_1$& $M_2$&  $\mu$&  $m_{\tilde{Q}}$
&  $m_{\tilde{l}}$&  $m_{\tilde{e}}$& 
$m_{3/2}/R$&$\alpha$\cr 
\hline
0.1&  1055& 184& 350& 700& 957&  435& 354&  450&   $0.26\times 10^{-2}$\cr 
30&1045& 436& 536& 607& 913&  531& 476&  631& 0.92\cr 
40&1038& 573& 653& 545& 879&  578& 541&  729&  1.24\cr 
45&1034& 657& 717& 499& 852&  604& 576&  790&  1.41\cr 
55&1020& 882& 892& 339& 765&  671& 675&  951&  1.74\cr   
\hline
\end{tabular}
\caption{Example of ISAJET solution of the low energy 
mass spectrum for  model A, for different value 
of $R$. Here we fix the low energy gluino mass roughly constant 
by choosing $M_3$(GUT)=450~GeV and $\tan\beta=10$. 
 All mass parameters are 
given in GeV. }
\label{table:massA}
\end{center}

\end{table}

As can be seen in Table \ref{table:massA},
the higgsino mass parameter $\mu$ decreases  as $\alpha$ increases. 
The  $\mu$ is determined by 
solving the minimization condition of one loop higgs effective 
potential.  In minimal supergravity model ($R=0$ limit), 
the higgs soft mass  square at weak scale 
is driven to   large negative value by stop/top loop, while 
the correction from gaugino loops is small  because 
$M_1, M_2\ll M_3$.  The $\mu$ parameter  
 is chosen to compensate the negative value to 
get correct  gauge symmetry 
breaking, therefore $\mu$ can 
be as large as the stop mass in the minimal supergravity model. 
When 
$\alpha$ is large, $M_1, M_2\gg M_3$ at the GUT scale. The 
higgs soft mass parameters get large positive contribution from 
the gaugino masses. 
Top/stop loop effect is largely compensated by the 
gaugino corrections at the weak scale. As a result, 
$M_1\sim \mu$ for $R\sim37$,  and 
$\mu\ll M_1, M_2$ for larger value of $R$.
Although $M_1$ and $M_2$ get closer to $M_3$ for 
$R\gg 37$, the 
mass splitting among SUSY particles increases again. 

We have seen in the model A  that 
decrease of the $\mu$ parameter limits the mass 
degeneracy. In our study, we are interested in the model point where 
the mass splitting among the SUSY particles are 
the smallest.  
The $\mu$ parameter would be largest when 
the GUT scale value of the higgs soft SUSY breaking 
mass is the smallest.  On the other hand, sfermion 
masses at GUT scale must be large to 
avoid the $\tilde{\tau}_1$ or $\tilde{t}_1$ LSP. 
 We therefore also consider the model 
B, where $n_{H_u}=n_{H_d}=1$ and  $n_{\rm matter}=0$ 
as another  example of the MMAM model. 

The mass spectrums of the models A and B for 
$M_{3}{\rm (GUT)}=450$~GeV and $\tan\beta=10$ 
are compared in 
Table \ref{table:massAB}. 
The mass difference between squark (gluino) and LSP 
hits minimum when $M_1\sim \mu$ at $R=R_c$.
 $R_c\sim 40$ for the model A and 55 for the model B, where 
$m_{\tilde{\chi}^0_1}/m_{\tilde{q}}=  0.55$ and $0.70$, 
respectively. 
As expected,  the model set B has more degenerated mass 
spectrum  at $R_c$, 
because the $\mu$ parameter is larger for the same 
gaugino masses. 
\begin{table}[thb]
\begin{center}
\begin{tabular}{|c|cc|cc|}
\hline
& set A & & set B& \cr
\hline
R& $m_{\tilde{u}_L} (m_{\tilde{g}}) m_{\tilde{\chi}^0_1}$ & 
$2p_{\rm CM} $& $m_{\tilde{u}_L} (m_{\tilde{g}}) m_{\tilde{\chi}^0_1}$ & 
$2p_{\rm CM}$\cr
\hline
 0& 995 (1055) 182 &961&  1041 (1061) 189 & 1007 \cr
10&  986 (1053) 246 &924&  1043 (1061) 248 & 984\cr
20& 973 (1049) 326  & 793 & 1044 (1060) 330&940\cr
30& 951 (1045) 426 & 726&  1045 (1060) 434 & 865 \cr
40 & 916 (1038) 507 & 635 & 1044 (1059) 569 &733 \cr
50& 854 (1027) 426 & 641& 1038 (1057) 713& 548 \cr
55& 803 (1021) 335& 663& 1033 (1056) 721 & 529 \cr
60& no EWSB& & 1023 (1055) 700 & 543 
\cr\hline
\end{tabular}
\caption{The squark, gluino, and the lightest neutralino masses 
in model A and B. $M_3{\rm (GUT)} =450$~GeV and $\tan\beta=10$. 
}
\label{table:massAB}
\end{center}
\end{table}


In the following sections, we discuss the 
discovery potential in the mass degenerated 
SUSY models. Phenomenologically important 
parameter for  the hadron collider 
is the typical mass scale  of the event. This can 
be expressed by the energy of the jet from $\tilde{q}
\rightarrow \tilde{\chi}^0_1 q$ decay in the $\tilde{q}$ 
rest frame, which is  expressed as 
\begin{equation}
2p_{\rm CM}=(m_{\tilde{q}}^2-m_{\tilde{\chi}^0_1}^2)/
m_{\tilde{q}}
\end{equation}
Because sparticles are pair produced, 
$2p_{\rm CM}$ is the typical order of the effective transverse 
mass $M_{\rm eff}$ of the SUSY production process. 
We list the value of $2p_{\rm CM}$ for the model points
in Table~\ref{table:massAB}. 
$M_{\rm eff}\propto 2p_{\rm CM}$ can be reduced 
by a factor of $2/3$ to $1/2$ for these models.  In the following 
sections, we find that discovery of SUSY particles will be 
non-trivial  in this region. It is worth noting that 
 gaugino and higgsino are highly mixed when   $p_{\rm CM}$ is minimum. 
 In this case, relatively  large nucleon-LSP scattering 
 cross section and smaller dark matter 
 density are expected, possibly consistent with
cosmological constraints.

Another phenomenologically important 
aspect  in terms of collider physics is the 
branching ratios  of the gluino and squark into leptons. 
The leptons would be  produced from the neutralino 
or chargino decays arising from the squark decays. 
For the model A, the decay $\tilde{q}\rightarrow \tilde{\chi}_i$
$(\sim \widetilde{W}, \widetilde{B})$ 
channels are  open unless 
$R>50$.  The chargino and neutralino may  decay
 into (s)leptons.  
Especially, 
if the decay channel $\tilde{q}\rightarrow 
\tilde{\chi}^0_i\rightarrow \tilde{l}
\  (l=e,\mu)\rightarrow \tilde{\chi}^0_1$ is open, 
the large branching ratio into the golden mode
$\tilde{q}\rightarrow \tilde{\chi} qll$ is expected. 
For example, ${\rm Br}(\tilde{\chi}^+\rightarrow \tilde{\tau}\nu)\sim$ 80\%
and ${\rm  Br}(\tilde{\chi}^0_2\rightarrow \mu^+\tilde{\mu}^-) \sim $14\%
for the model A, $M_3{\rm (GUT)}=450$~GeV, $\tan\beta=10$, and $R=37$.  
For the model A, $M_2>m_{\tilde{l}}$  if $R>20$, namely,   the decay channel 
is open especially in  the degenerated region.   
For the model B, the  squark and slepton masses 
are so heavy that the decay into slepton is 
always closed.

\section{Monte Carlo simulation and reconstruction}
Here 
we describe our event simulation method used 
in 
the next section. 
As explained 
earlier, SUSY mass spectrum 
is calculated by ISAJET~\cite{ISAJET} which is interfaced
to the HERWIG~\cite{HERWIG} event generator
using ISAWIG~\cite{ISAWIG}.  
HERWIG generates hard processes,  
takes care of initial and final state radiations, and  fragments 
partons into hadrons.  

To estimate event distributions to be  measured at
LHC detectors, we smear particle energies, 
identify isolated leptons, and reconstruct jets. 
We independently developed a fast detector simulation program 
\footnote{Recently,  a fine  
event simulator PGS (pretty good simulation) \cite{PGS}
is also available. },  which takes the following steps;%
\begin{enumerate}
\item  {\bf Finding isolated leptons:} If a lepton ($e$ or $\mu$)  
with $E_{T}>10$~GeV and 
$\vert\eta\vert<2.5$ is found in an event record,  
we take a cone 
with a size $\Delta R=\sqrt{(\Delta\eta)^2+(\Delta\phi)^2}=0.2$ 
around the lepton. If the sum of 
$E_T$ of the particles  
in the cone (except the lepton at the center of the cone)  
is less than 
10~GeV, we regard it as 
an isolated lepton candidate.  
After the jet reconstruction described below,  
isolated  lepton candidates  with ${\rm min}(\Delta R(lj))>0.4$
are accepted as isolated leptons. 
\item  {\bf Reconstruction of jets}.  
We adopt a simple jet finding algorithm  PYCELL, 
a subroutine  in the PYTHIA package,  with 
minor modification on its treatment of leptons and 
energy smearing.  Namely, 
\begin{enumerate}
\item We remove  isolated lepton 
candidates for the jet reconstruction.
\item  Particle energies are  measured by the 
electromagnetic and hadronic calorimeters. 
We assume that the number of calorimeter cells in $\eta$ direction 
is 50 within  $\vert \eta \vert <3$ and 
50 in  $\phi$ direction, respectively.  
Transverse energy deposit $E_T$  in a 
cell is summed for electrons, photons, and hadrons  as 
\begin{eqnarray}
E_T({\rm EM})
&=&\sum_{\rm electrons} E_T+ \sum_{\rm photons} E_T\cr
E_T({\rm had})
&=&\sum_{\rm hadrons} E_T,
\end{eqnarray}
which correspond to the hits in the electromagnetic 
and hadronic calorimeter cells,
respectively. 
\item Transverse energy deposit in  each cell is smeared by Gaussian 
energy resolutions 
$\sigma_{E_T}({\rm EM})=0.1 \sqrt{E_T({\rm EM})[GeV]}
$ and $\sigma_{E_T}({\rm had})=0.5\sqrt{E_T({\rm had})[GeV]}$.  
After the energy smearing,  we regard sum of the 
smeared $E_T({\rm EM})$ and $E_T({\rm had})$  
as the measured energy deposit in the cell. 
\item 
We take the  highest energy 
cell as the initiator and sum the energy deposits in the cell 
within $\Delta R<0.4$. We accept the cluster as a jet
if $\sum_{\rm cluster}E_{T} >10$~ GeV.  We repeat this procedure
after  removing the cells which have been 
already used for the jet reconstruction.
\end{enumerate}
\end{enumerate}

The reconstruction process (1 and 2) 
 is similar to  the fast simulation codes 
used by the ATLAS and CMS groups. 
Note that we do not smear energy of 
isolated leptons and assume 
they are identified correctly,  because  the energy 
resolution  and particle identification are  
excellent for leptons in both of the LHC detectors.  
We smear photon and non-isolated 
electrons energies  with better resolution 
than that for hadrons, however  we do not assume 
fine grained electromagnetic calorimeter in our simulation 
for simplicity. 

Our simulation does not take care of other detector effects, 
such 
as misidentification of leptons and hadrons, non-uniformity 
of detector responses (cracks), non-Gaussian smearing 
in the energy measurements. 
The simulation, however,  reproduces the 
signal distributions of past simulation studies reasonably well.  

\begin{figure}
\begin{center}
\includegraphics[width=6cm, angle=90]{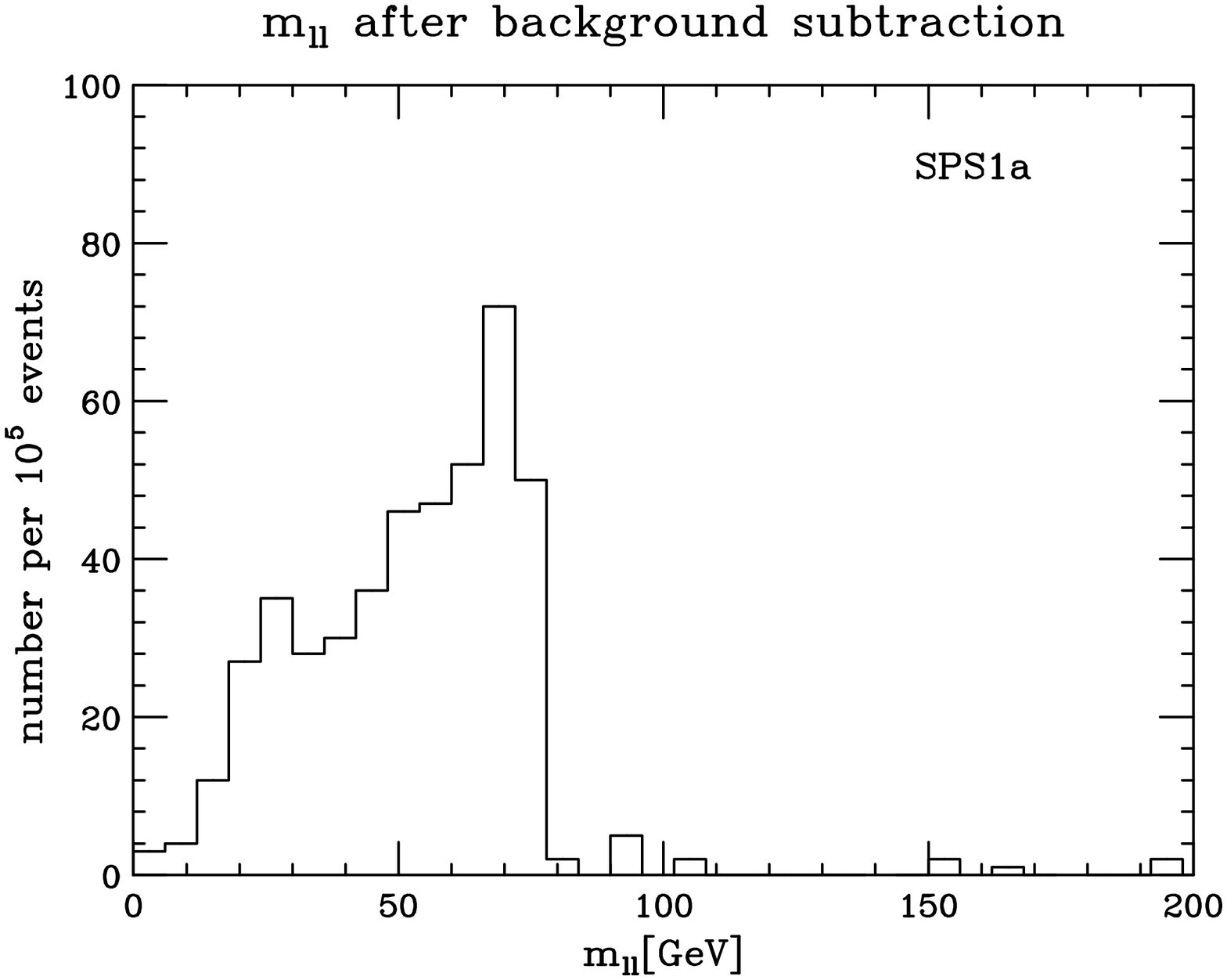}
\includegraphics[width=6cm, angle=90]{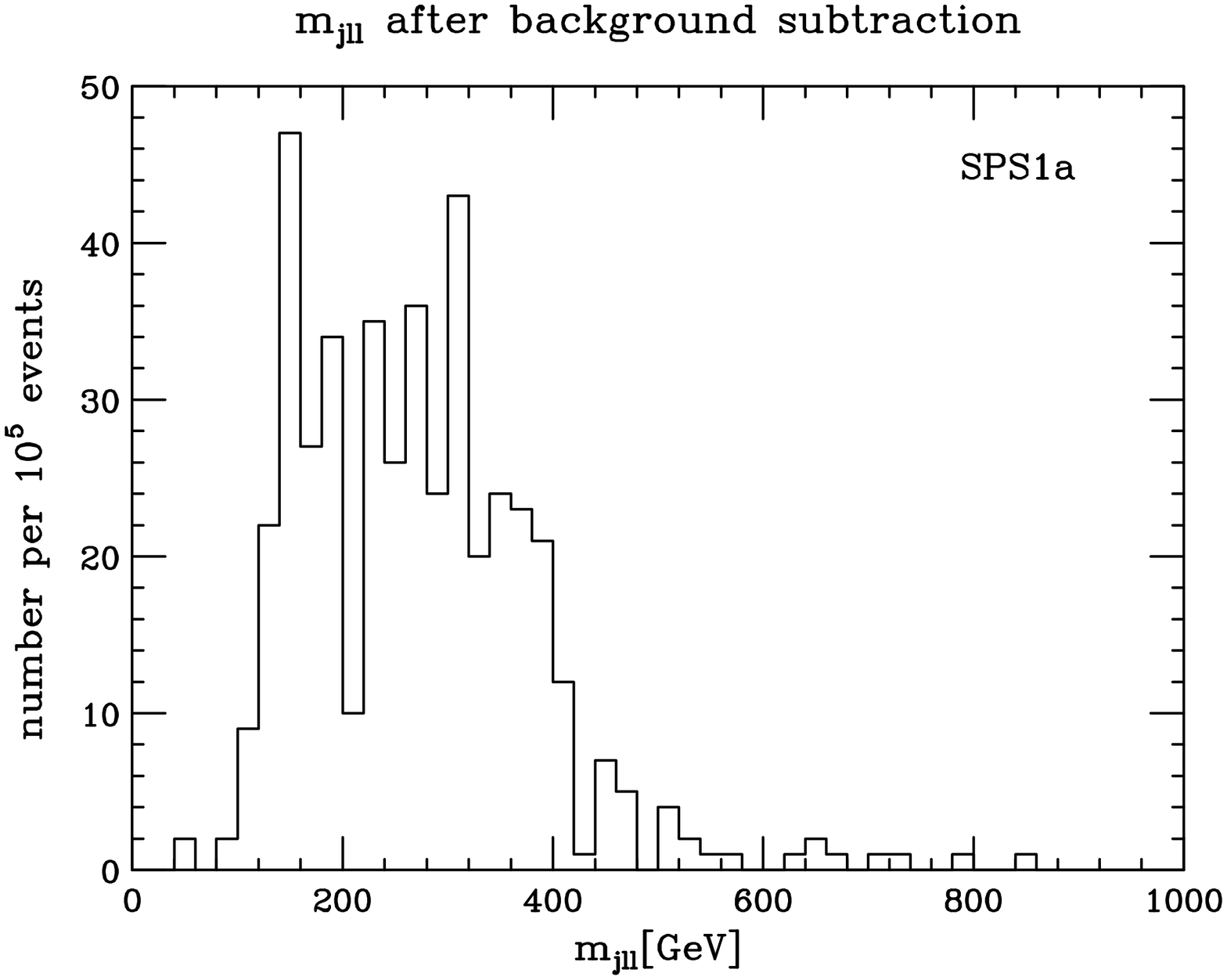}
\caption{ The dilepton invariant mass distribution $m_{ll}=m_{ee}+m_{\mu\mu}-m_{e\mu}$ (left) and $m_{jll}$(right) 
distribution at the  point SPS1a by our fast simulation program. We 
produce  $10^5$ 
SUSY events for the plot, corresponding to $\int dt {\cal L}=1.8$ fb$^{-1}$.
After the cuts and background subtractions, 445 dilepton events remain.  }
\label{fig:msps1a}
\end{center}
\end{figure}

To check the validity 
of our independently developed simulation program, 
we compare published 
distributions of SUSY signals  using the ATLFAST simulator~\cite{ATLFAST}
with our simulation results. 
Figure~\ref{fig:msps1a} 
show distributions of 
the  same flavor and opposite sign dilepton invariant mass
($m_{ll}$) and the jet-lepton-antilepton 
invariant mass ($m_{j_1ll}$) at 
SPS1a~\cite{snowmass}, 
where $j_1$  is one of the two highest $p_T$ 
jets with $m_{j_1 ll}<m_{j_2 ll}$. 
The distributions are important 
to reconstruct the cascade decay 
$\tilde{q} \rightarrow \tilde{\chi}^0_i
\rightarrow\tilde{l} \rightarrow\tilde{\chi}^0_1$. 
Contribution of background events with
accidental two leptons can be subtracted using the distributions  
of $e^\pm \mu^{\mp}$ events. 
The plots are produced 
by applying the same cuts as used in  \cite{LHCLC}; 
\begin{itemize}
\item $M_{\rm eff}> 600$~GeV.
\item at least 4 jets with $p_{T1}>150$~GeV, $p_{T2}>100$~GeV, and 
$p_{T3,4}>50$~GeV.
\item two and only two leptons 
with $p_{T}(l_1)>20$~GeV and $p_{T}(l_2)>10$~GeV. 
\item $S_T>0.2$
\item $\ETmiss >{\rm max}(0.2M_{\rm eff})$. 
\end{itemize} 
We calculate the sphericity and $\ETmiss$, 
based on the smeared energy deposits to the cells in $\vert\eta\vert<3$ 
in addition to the momentums of  isolated lepton candidates. 

We find that accepted number of events is consistent with 
the previous simulation studies. 
In Figure~\ref{fig:msps1a} 
we show the $m_{ll}$ and $m_{jll}$ distribution for  
$10^5$ SUSY events (corresponding to $\int dt {\cal L}=$ 1.8~fb$^{-1}$). 
The edge and the end point appear at correct positions. 
The $m_{ll}$ distribution 
has a sharper edge compared with that in  \cite{LHCLC},  because 
we do not smear lepton energies in our simulation. 
We find  that 445 events remain after the cuts and the background 
subtraction. In the report~\cite{LHCLC} roughly $2.5\times 10^4$ 
events remain after the cuts for $\int dt {\cal L} =$ 100~fb$^{-1}$. 
The acceptances of the two simulations  therefore 
agree within 1$\sigma$.

\section{Discovery of the degenerate SUSY model}
\subsection{Mass degeneracy and signal distributions}
In this sub-section, we study the $M_{\rm eff}$ distributions at
the SUSY model points  with degenerated mass spectrum. 

\begin{figure}
\begin{center}
\includegraphics[width=6cm, angle=90]{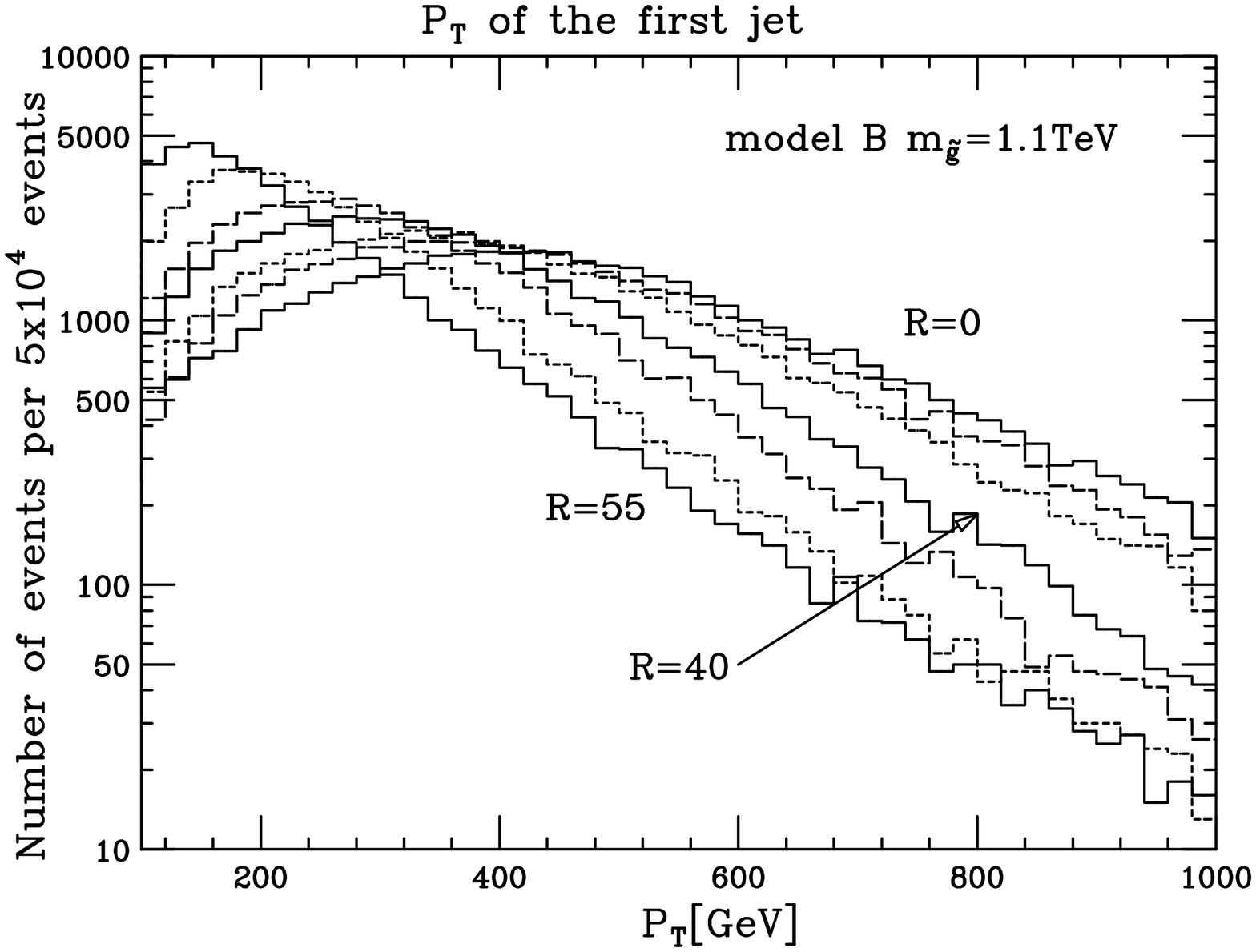}
\includegraphics[width=6cm, angle=90]{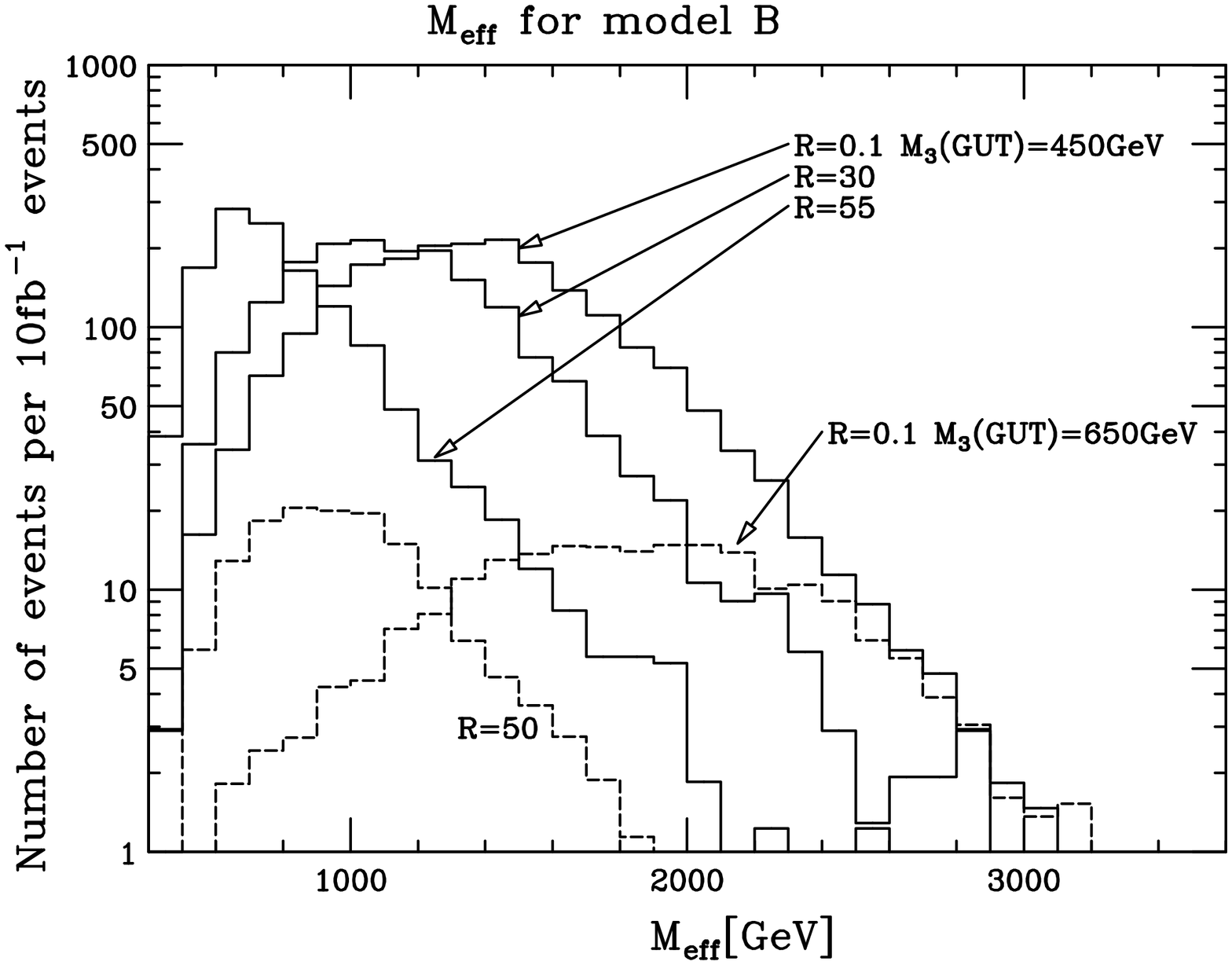}
\caption{ Left)  The $p_T$ distributions of the 
highest $p_T$ jet for model 
 B,  $M_3{\rm (GUT)}=450$~GeV and $\tan\beta=10$.  
50000 events are generated for each 
points($R=0,20,30,40,45,50,55$ from right 
to the left). Right)  $M_{\rm eff}$ distributions 
for $10{\rm fb }^{-1}$  for model B with $M_3{\rm (GUT)}=450$~GeV 
$R=0, 30, 55$ (solid histograms from right to left) and 
$M_3{\rm (GUT)}=650$~GeV $R=0, 50$(dashed histograms 
from right to 
left)  respectively. }
\label{fig:ptmeffB}
\end{center}
\end{figure}

We first show the $p_T$ distribution of the first jet 
in the left panel of Figure \ref{fig:ptmeffB}. 
Here we take the model B, $M_3{\rm (GUT)}=450$~GeV, 
$\tan\beta=10$ and vary degeneracy 
by changing $R$ from 0 to 55. 
As $R$ increases, sparticle mass difference gets smaller, 
resulting in softer
$p_T$ distribution.  
The peak positions are  positively correlated with 
$p_{\rm CM}$ which  ranges from $504$~GeV to $265$~GeV  when 
we change $R$ from $0$ to $55$. 
From the figure, we can read that the acceptance 
of SUSY events depends on $R$. For example, 
events with  $p_T<160$~GeV is only 16\% for the point with $R=0$ but 
it is 35\% for the point with $R=55$.  This means that 
about one third 
of the events are rejected by the requirement 
on the $p_T$ of the first jet  in the previous section for the latter 
point. 
 
The $M_{\rm eff}$ distribution shows a similar 
behavior.  In Figure~\ref{fig:ptmeffB} (right), 
we show the $M_{\rm eff}$ distributions for the model B,
with  
$M_3{\rm (GUT)}=450$~GeV (solid histograms) 
and $M_3{\rm (GUT)}=650$~GeV (dashed histograms), respectively. 
We apply the cut (A) and also require 
one  hard isolated  lepton  with  $p_T>20$~GeV and 
$\vert \eta\vert<2.5$.

Here we compare the distributions with 
$R=0.1$ (MSUGRA like)  and  $R=30$ to $55$ (degenerated).  
It should be noted that, while the mass spectrums are considerably 
different,  the power low of the distributions 
beyond their peaks are roughly the same.  
These high $p_T$ events are originated from the 
collisions with its center of mass energy $\sqrt{s}$  much higher 
than the squark and gluino masses.
Therefore 
the power low of the distributions  only depends on the 
dominant luminosity function.

The peak position of the $M_{\rm eff}$ distribution has a direct correlation 
with the produced sparticle mass.
In \cite{TOVEY} it is 
found that the correlation between the SUSY mass scale defined as 
\begin{eqnarray}
M^{\rm eff}_{\rm SUSY}&=&\left(
M_{\rm SUSY}-\frac{M^2_{\chi}}{M_{\rm SUSY}}\right)
\end{eqnarray}
where 
$M_{\chi}$ is the LSP mass and
$M_{\rm SUSY}
=({\sum \sigma_{i} m_i})/({\sum \sigma_{i}})$.
The peak position of 
$M_{\rm eff}$ is linear  for MSUGRA and Gauge mediation 
model. We also find that the linear relation holds for the signal 
distribution. We do not provide the fit results here, because 
the existence of the standard model background is very important for this 
model, as will be discussed in the following subsections
\footnote{The paper \cite{TOVEY} also found 
the linear relation does not hold in   general SUSY model. 
These are corresponding to the points where 
the dominant contribution to 
the total SUSY production cross section comes from lighter 
sparticles such as chargino, neutralino, and sleptons, 
which  do not contribute to 
the 4 jet + missing $E_T$ signals. }.

\begin{figure}{t}
\begin{center}
\includegraphics[width=6cm, angle=-90]{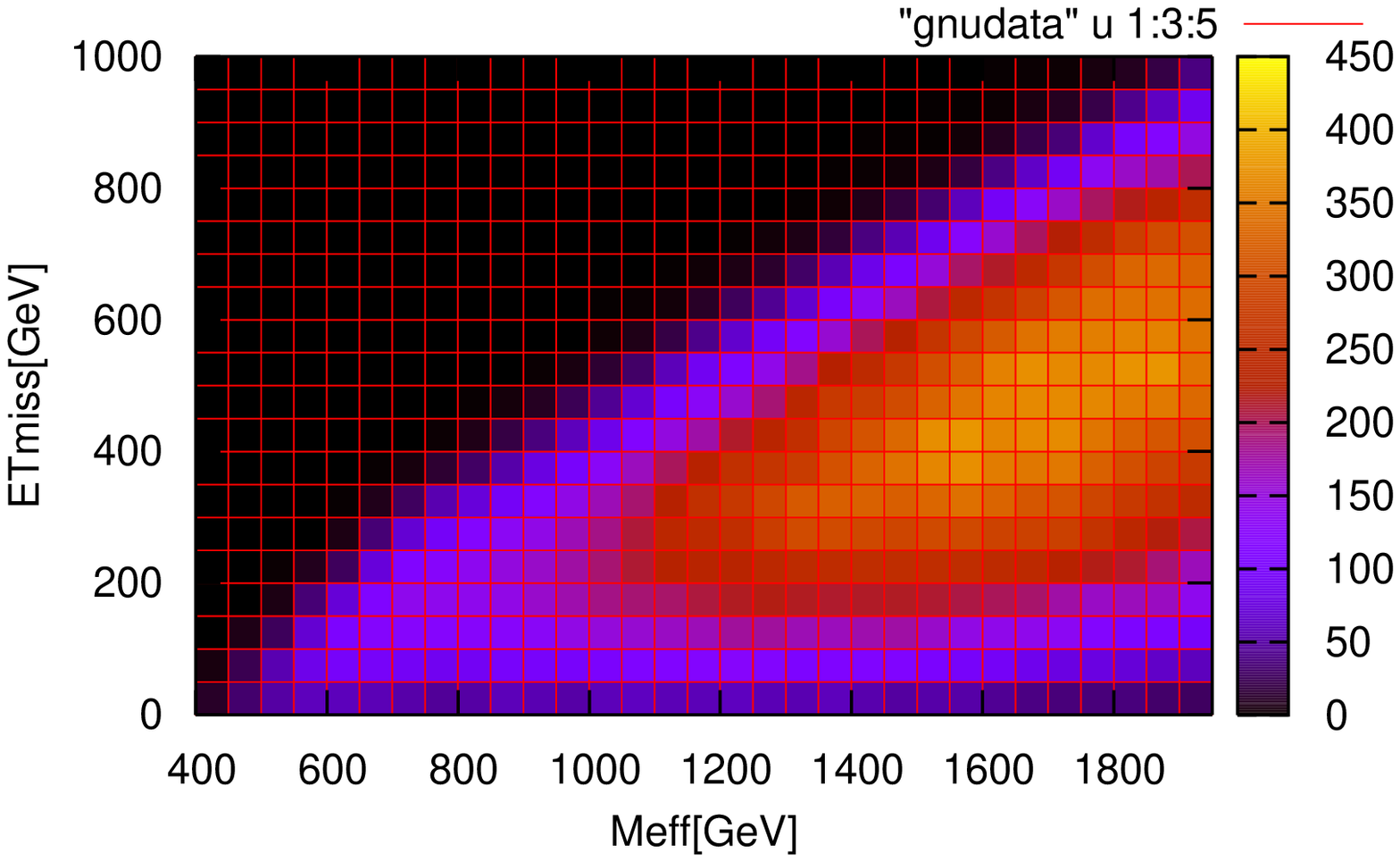}
\includegraphics[width=6cm, angle=-90]{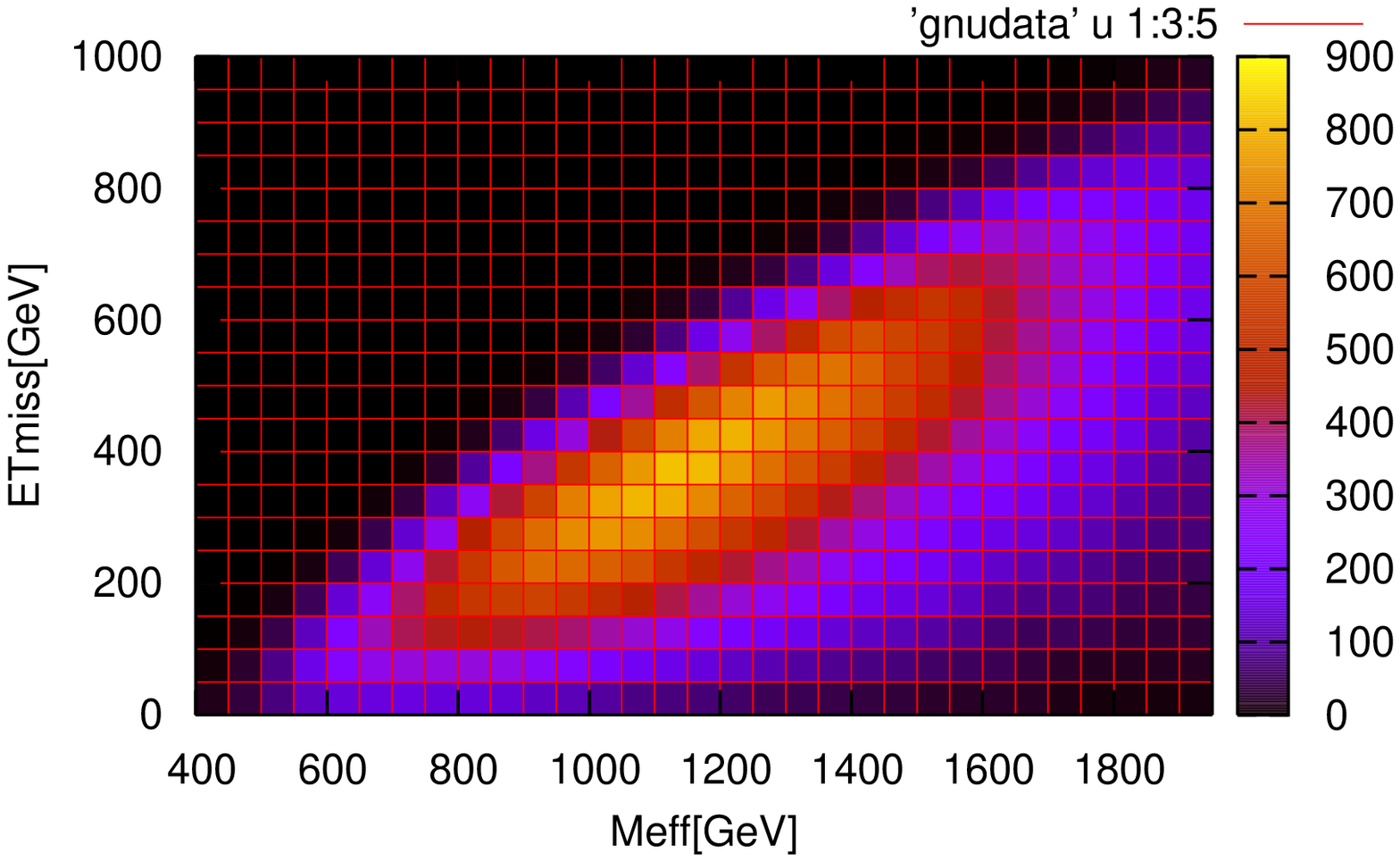}\\
\includegraphics[width=6cm, angle=-90]{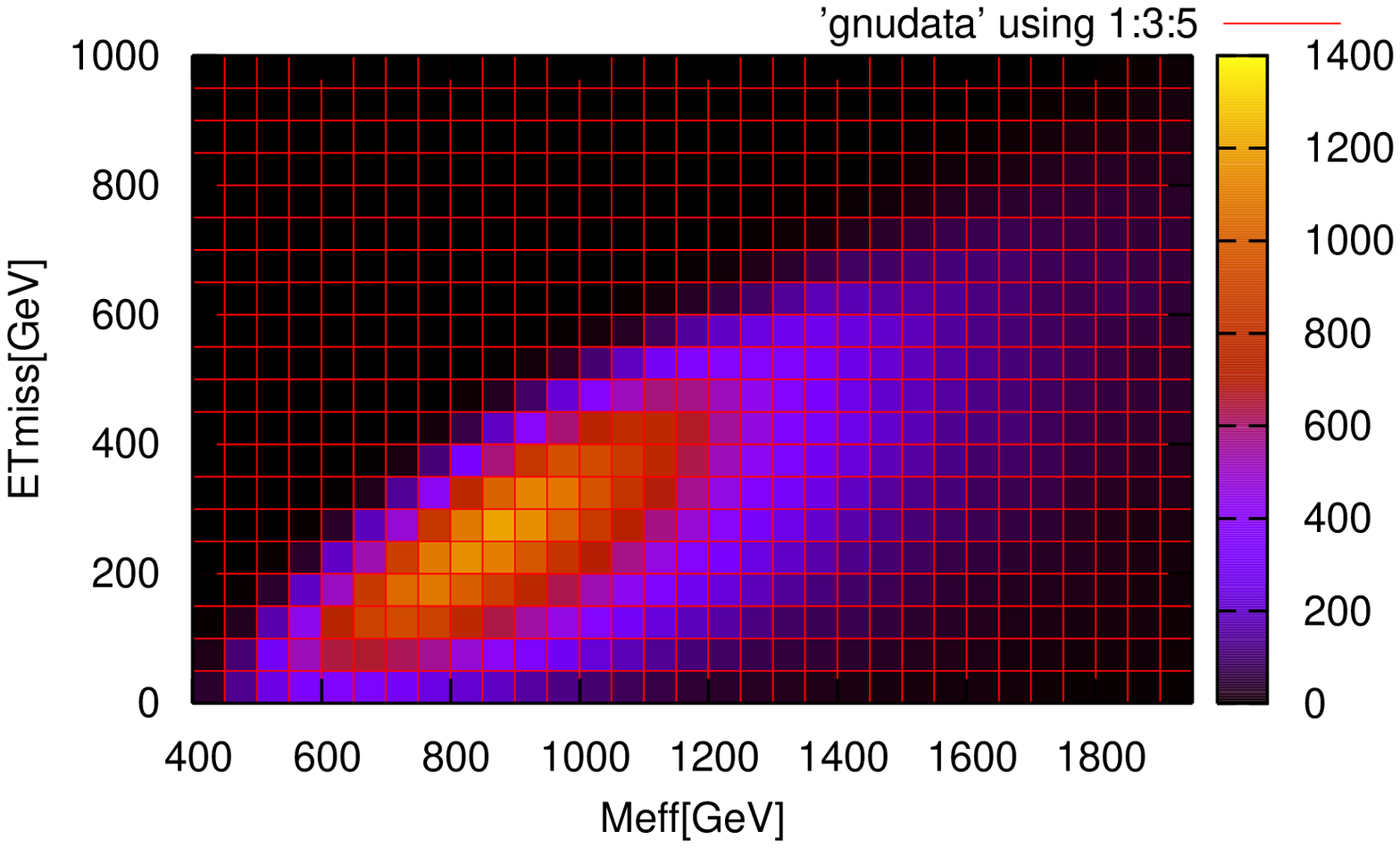}
\caption{ $M_{\rm eff}$  vs $\ETmiss$ distribution for 
the model B, with $M_3{\rm (GUT)}=650$~GeV, $\tan\beta=10$ and 
$R=0.1$ (top left), $R=40$ (top right) and $R=50$ (bottom) respectively.} 
\label{fig:etmissmeffB}
\end{center}
\end{figure}

We now turn into the relation between 
$\ETmiss$ and $M_{\rm eff}$ for degenerated 
points.  
In Figure~\ref{fig:etmissmeffB}
we show the distribution for the model B
with $M_3{({\rm GUT})}=650$~GeV,  $\tan\beta=10$ and $R=0.1, 40, 50$,
respectively.    The $x$- and $y$- axes are $M_{\rm eff}$ and
$\ETmiss$, respectively,  in Figure~\ref{fig:etmissmeffB}.   
We see that  the $\ETmiss$  takes a significant 
part of $M_{\rm eff}$ for degenerated points $R=40$, and $50$ if $M_{\rm eff}$ 
is higher, while, 
for $R=0.1$ the $\ETmiss$ for a fixed $M_{\rm eff}$ is broader in the 
plot.

This distribution can be understood as follows. Suppose 
we have events with  two uncorrelated jets with energy 
$E_{\rm jet}=p_{\rm CM}$.   The dominant part of the 
cross section is squark/gluino pair production 
near the threshold for our case, so 
if they decay directly into two jets and two LSPs,  the event kinematics 
 are indeed of this type up to the boost to the beam direction. 
We now calculate 
missing energy $E_{\rm miss}$ and effective energy $M{\rm (eff)}$ of the events. 
The momentum of the two jets and the  missing momentum  
can be expressed as follows, 
\begin{eqnarray}
p_1&=&(p_{\rm CM}, 0, 0, p_{\rm CM}) \cr
p_2&=&(p_{\rm CM},  p_{\rm CM}\sin\theta, 0, p_{\rm CM}\cos\theta) \cr
p_{\rm miss}&=& (E_{\rm miss}, -p_{\rm CM}\sin\theta, 0, -p_{\rm CM}(1+\cos\theta) ) 
\end{eqnarray}
where  we take the direction of the $p_1$ momentum  as $z$-axis and 
events are in $x$-$z$ plane. 
Then the missing energy $E_{\rm miss}$ and effective  
mass $M{\rm (eff)}$ are defined as 
\begin{eqnarray}
E_{\rm miss}&=& p_{\rm CM}\sqrt{2+2\cos\theta},\cr
M{\rm (eff)}&=&2p_{\rm CM}+E_{\rm miss},
\end{eqnarray}
where  $M{\rm (eff)}$ ranges from $2p_{\rm CM}$ to $4p_{\rm CM}$. 

The relation between $\ETmiss$ and $M_{\rm eff}$ is somewhat 
similar to that of $E_{\rm miss}$ and $M{\rm (eff)}$ 
as can be seen in Figure~\ref{fig:etmissmeffB}. 
For each plot 
$\ETmiss$ tends to be small when 
$M_{\rm eff}\sim  2p_{\rm CM}$, where $2p_{\rm CM}=1.36, 0.95, 0.7$~TeV  
for $R=0.1, 40, 50$, respectively.  It increases  linearly as a function
of $M_{\rm eff}$ up 
to $M_{\rm eff}\sim 3p_{\rm CM}$, and
the number of events reduces  quickly beyond $M_{\rm eff}> 4p_{\rm CM}$ .   
The relation indeed should be  true when two  particle 
are pair produced and  each of them decays into a visible particle and 
an invisible particle at  hadron collider.  

On the other hand,  for the 
dominant background coming from  $W(Z)$+ $n$-jets or $t\bar{t}$ +$n$-jets, 
the $\ETmiss/M_{\rm eff}$ should be 
smaller than those of the signal.   
This is because  the SM events have many other jet activities which are 
not associated with the missing momentum productions.  
The background  
comes from highly boosted events, where $\ETmiss$ is typically a  fraction of 
the missing energy originated from high energy $W$, $Z$, and $t$. 
The canonical cut to reduce the SM background is  
$\ETmiss >0.2 M_{\rm eff}$. 
If the background distribution in  $M_{\rm eff}$ and $\ETmiss$
plane is significantly different from the signal distribution, one can 
improve the S/N by applying an improved cut 
in the plane rather than the simple  cuts 
$M_{\rm eff}>400$ GeV  
and $\ETmiss > 0.2 M_{\rm eff}$
described in section 1. 
Because there are no background distribution study  on the relation between 
$M_{\rm eff}$ and $\ETmiss$, we do not attempt to improve 
the cuts in following sections and use the cut 
given in section 1. 

\subsection{The zero lepton channel}
We now show the signal $M_{\rm eff}$ distributions for the 
degenerated model. 
We start our discussion  from the $M_{\rm eff }$ distributions of the 
SUSY events without  high $p_T$ leptons. This channel 
has been studied in \cite{TDR, HP, TOVEY} with an emphasis on 
the model-independent reconstruction of the SUSY 
scale, as discussed in the previous  subsection.

\begin{figure}
\begin{center}
\includegraphics[width=6cm,angle=90]{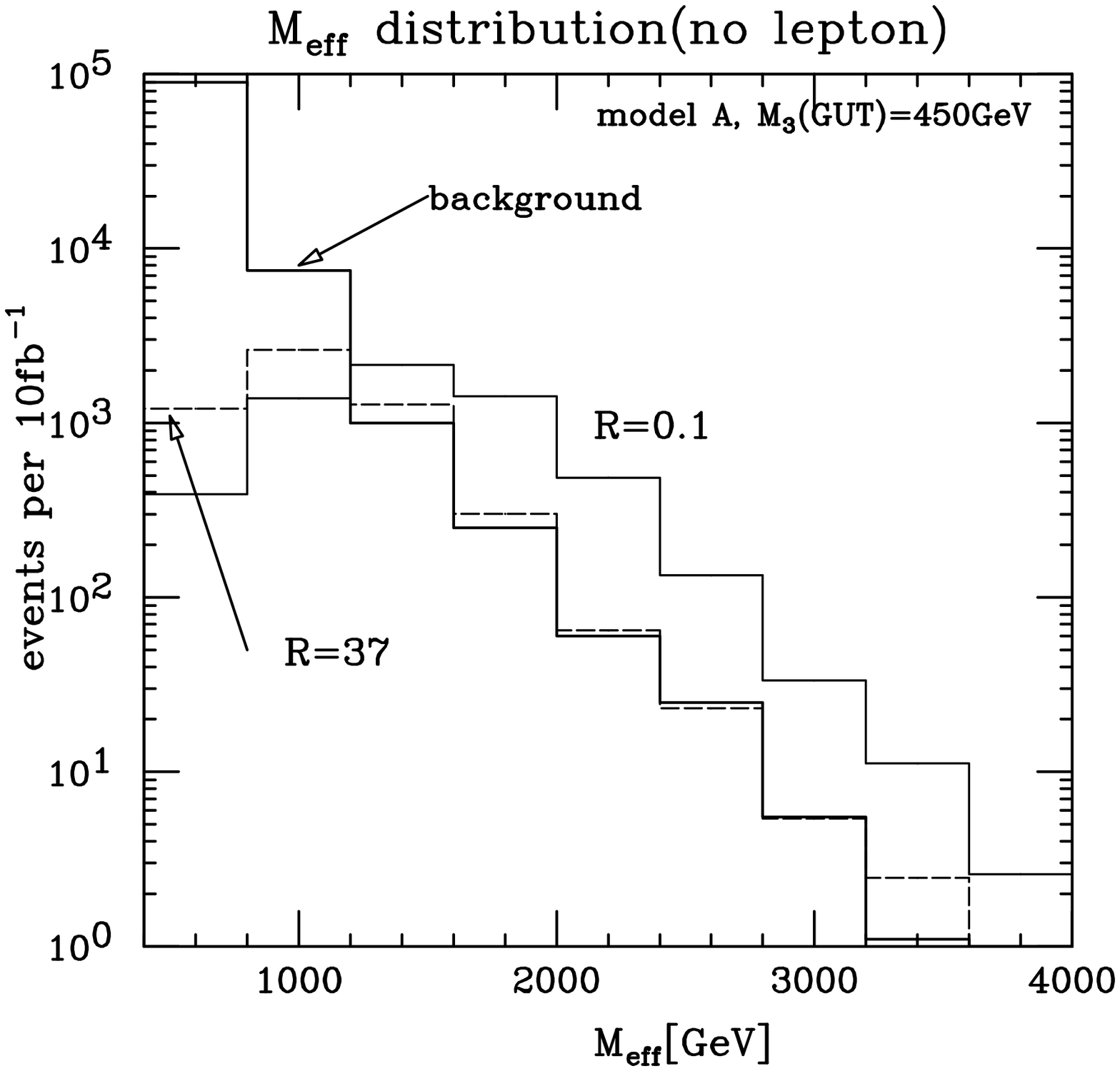}
\includegraphics[width=6cm,angle=90]{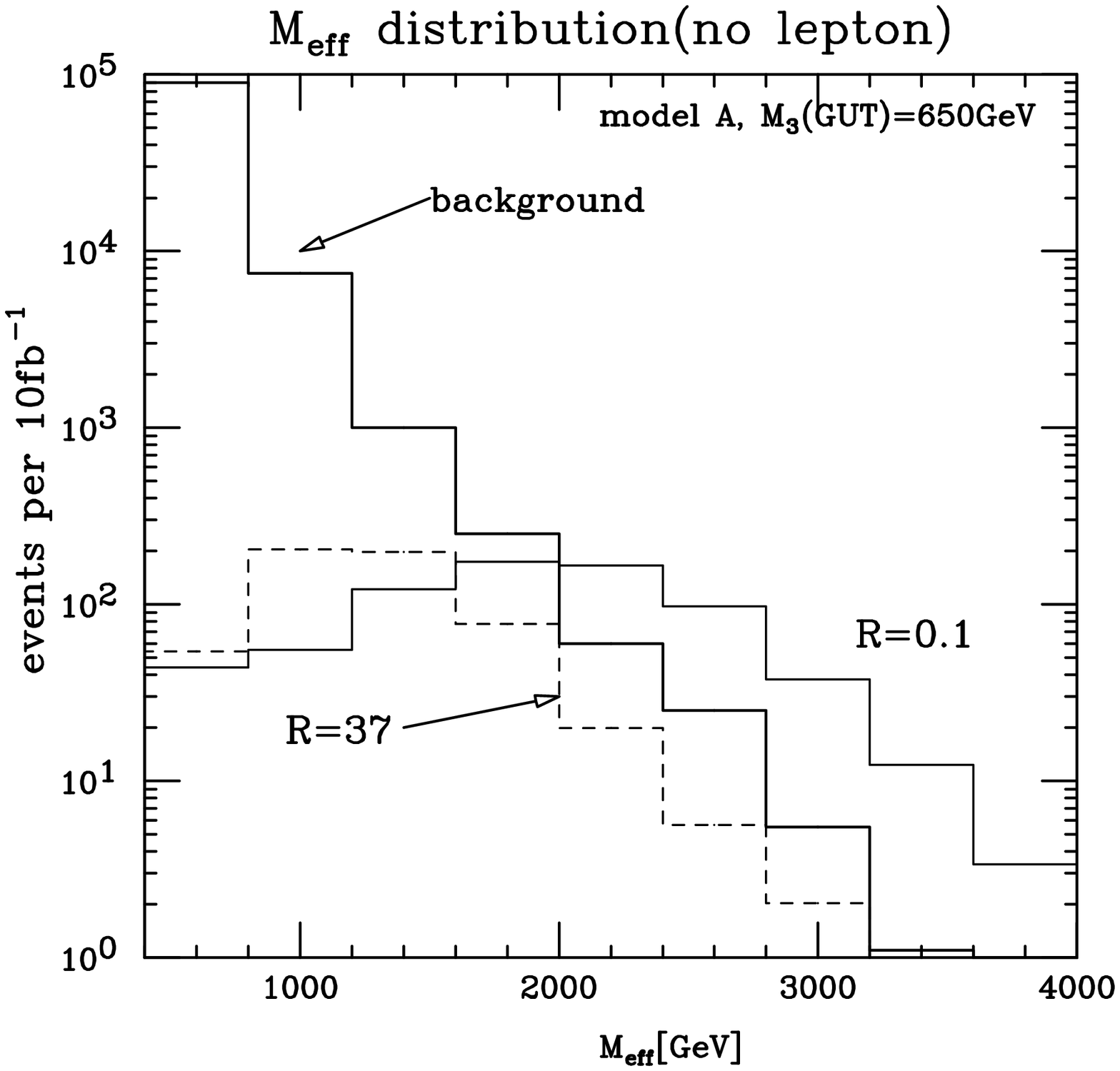}
\caption{The $M_{\rm eff}$ distributions 
for non-lepton sample.
The  thick solid histogram  is the background estimated in \cite{TDR}. 
Solid and dashed histograms 
 are  signal distributions 
 for $M_3{\rm (GUT)}=450$~GeV $R=0.1$ (MSUGRA),
  and $R=37$  (degenerate) 
 in the  left panel,  and  for $M_3{\rm (GUT)}=650$~GeV 
$R= 0.1$ (MSUGRA) and $R=38$ (degenerate) 
in the right panel, respectively. 
 All plots are normalized for $\int dt {\cal L}= 10$ fb$^{-1}$. }
\label{fig:meffA}
\end{center}
\end{figure}

In  Figure \ref{fig:meffA},  the signal distributions for the model A, 
with $M_3{\rm (GUT)}=450$~GeV and 650~GeV are plotted for 
$R\sim 0$ and $R\sim 37$.  The signal distributions
(solid and  dashed histograms)
are parallel to the background 
distribution (bars in the plots) beyond the peak of 
$M_{\rm eff}$ distribution. The difference 
appears only in the overall normalizations. 
As already discussed, this is because 
the signal distribution is determined  mainly by the  
luminosity function, once  $M_{\rm eff}\gg 2M_{\rm SUSY}$. 

The background shown in the plots  has a large uncertainty. 
The estimation has been made by the lowest order  Monte Carlo 
generator ISAJET.  The distribution is subjet to the scale 
uncertainty which is typically ${\cal O}$(30\%).  
We  also require at least 4 high $p_T$
 jets, while the number of final state parton involved in the tree 
level diagram is much less than 4. 
The number of 
additional jets is estimated by parton shower (PS) approximation. 
The PS approximation is good to  estimate 
distributions of jets collinear to the leading jets.
Recently, several groups have emphasized the importance 
of the matrix element (ME) correction 
to the SM background processes. 
The matrix elements of the diagram involving  $W, Z, t$ + $n$-partons 
are calculated  by the generator devoted for the process 
with multi jets~\cite{MANGANO, RAINWATER, Maltoni:2002qb}. 
Experimental groups are also working to 
take into account the corrections, 
for example, in the ATLAS group~\cite{TEV4LHC}, 
the backgrounds from $(\bar{t} t,W, Z)$+$n$-jets  are  calculated  
up to $n=(3,6, 6)$ respectively using ALPGEN~\cite{MANGANO}. 
After the inclusion of the ME corrections, the SM background is increased by 
a factor of 2$\sim$4 for $M_{\rm eff}>2$~TeV. Even in the 
highest bin in the plot, the expected background is above 1.  
This is caused by the increased high $p_T$ 
jets in the SM events.   The improved background estimation is still
preliminary and  
in the level of the lowest order, and subject of coupling scale uncertainty. 

We find that the signal rate is the same  order of the background rate 
for the  degenerated parameters. 
If the uncertainty of the 
overall normalization of the background is 100\%, no 
bound would be obtained from no-lepton channel by 
looking into $M_{\rm eff }$ distribution only. 
However, as we noted previously, the structure of the 
signal distribution may be visible over the background 
distribution in the $M_{\rm eff }$ and $\ETmiss$ plane. 

\subsection{One lepton channel}
\begin{figure}
\begin{center}
\includegraphics[width=4.5cm,angle=90]{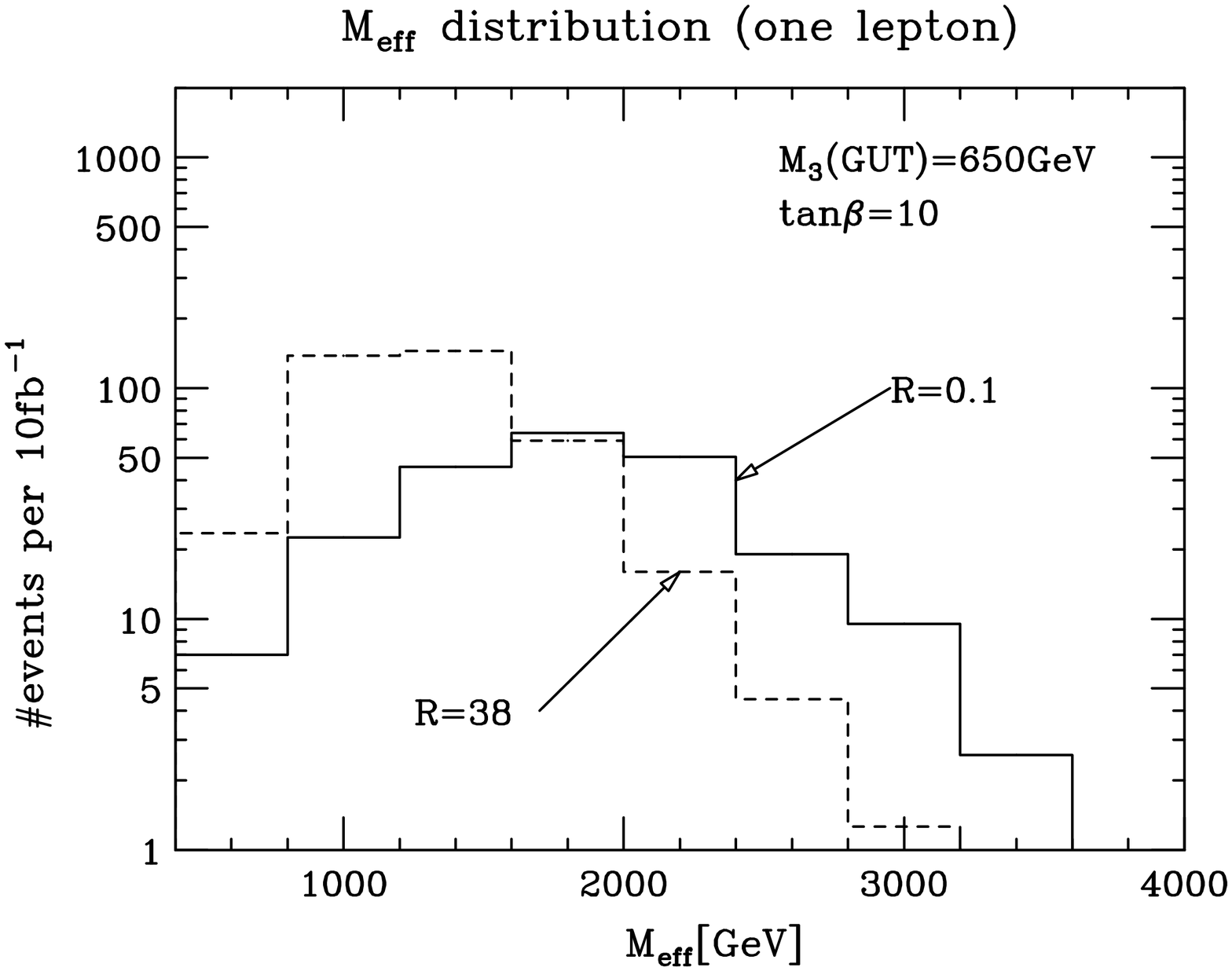}
\includegraphics[width=4.5cm,angle=90]{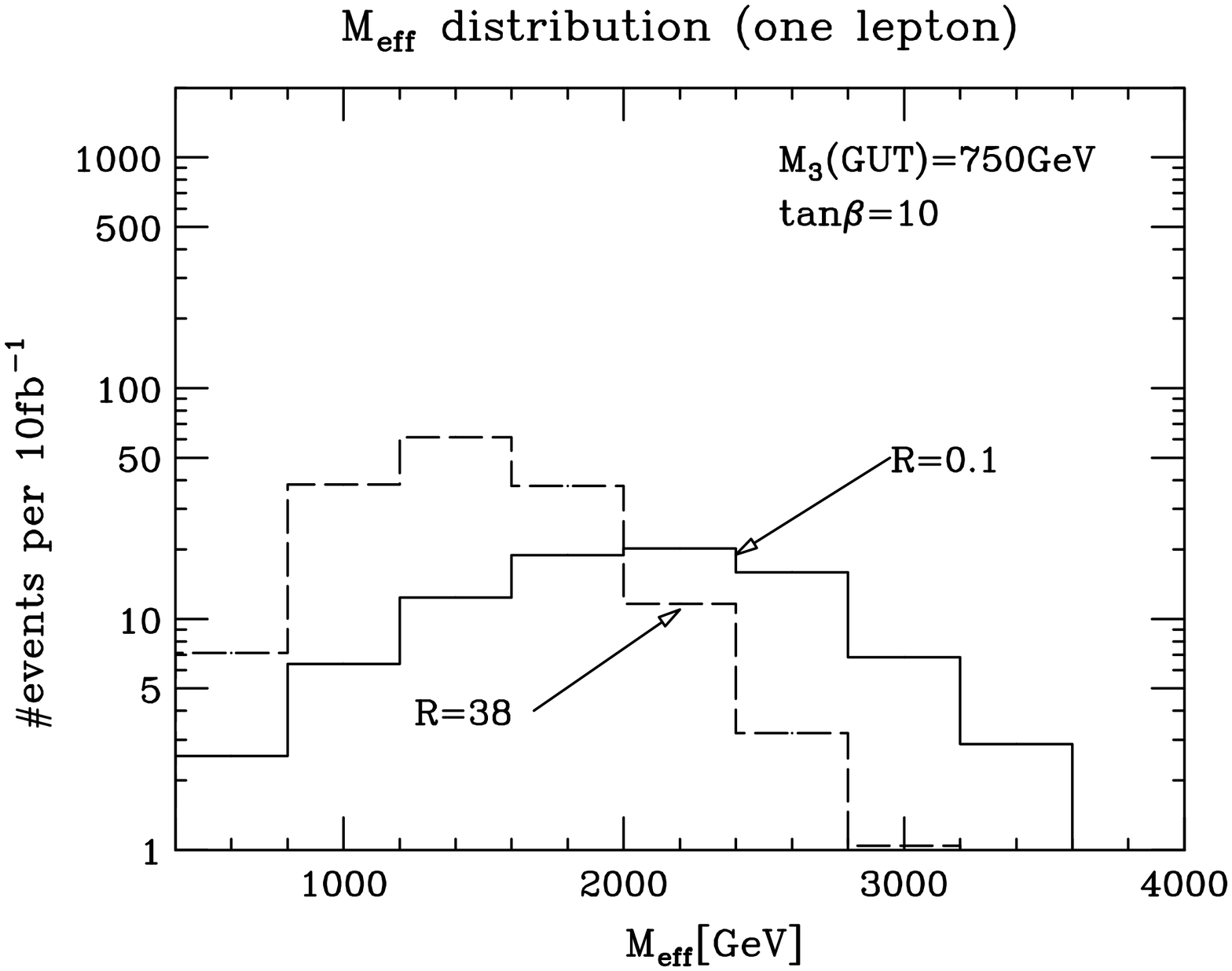}
\includegraphics[width=4.5cm,angle=90]{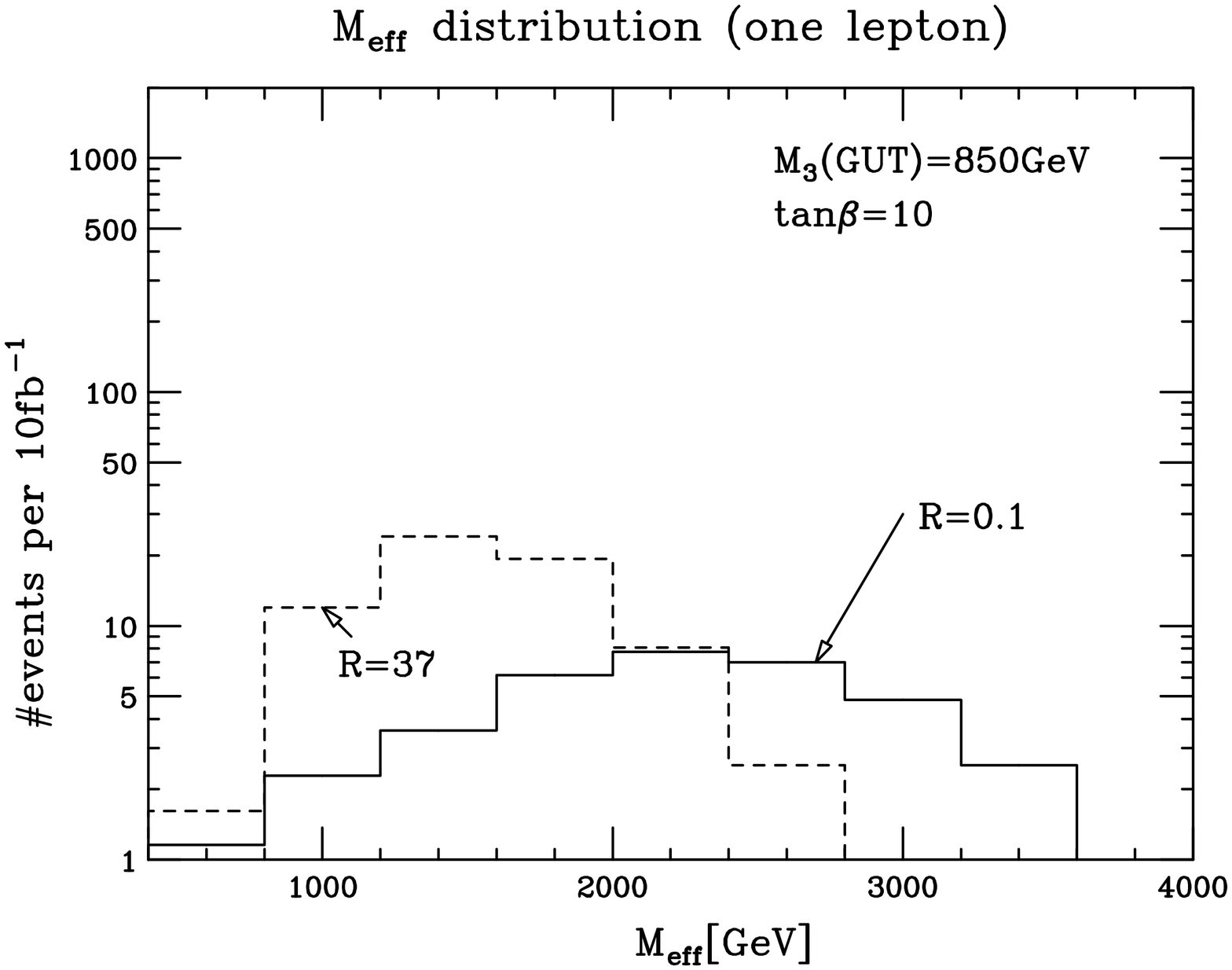}
\caption{The signal  $M_{\rm eff}$ distributions 
of one lepton events. 
The solid  and dashed histograms correspond to 
model A, $R\sim 0$ (MSUGRA-like) and $R\sim 37$
(degenerated mass spectrum) respectively. $M_3{\rm (GUT)}= 650$, 
 750, 850~GeV from left to right. All plots are normalized for 
 $\int dt L =10$fb$^{-1}$. }
\label{fig:mefflA}
\end{center}
\end{figure}

The significant fraction of SUSY production gives rise to events with 
isolated  leptons ($P_T >20$~GeV and $\vert\eta\vert <2.5$). 
We show the signal distribution for the model A in  Figure~\ref{fig:mefflA}.
Here, we do not show  the background distributions, because 
they are not given in previous literatures.  In \cite{TEV4LHC}, 
the distribution of the SM background is given, which 
is roughly 1/20 of the background of the zero lepton channel 
with a similar shape.  

\begin{figure}
\includegraphics[width=6cm,angle=90]{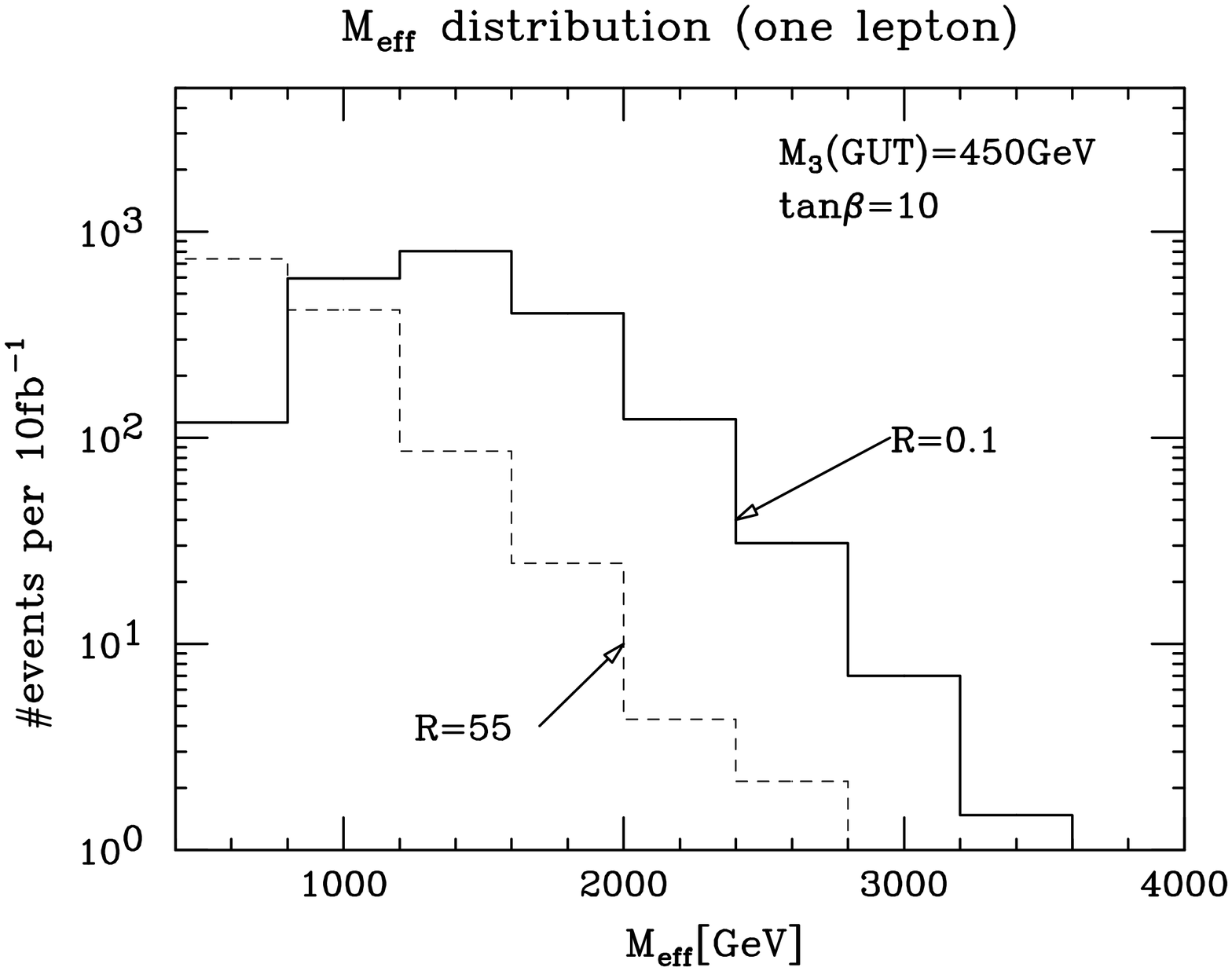}
\includegraphics[width=6cm, angle=90]{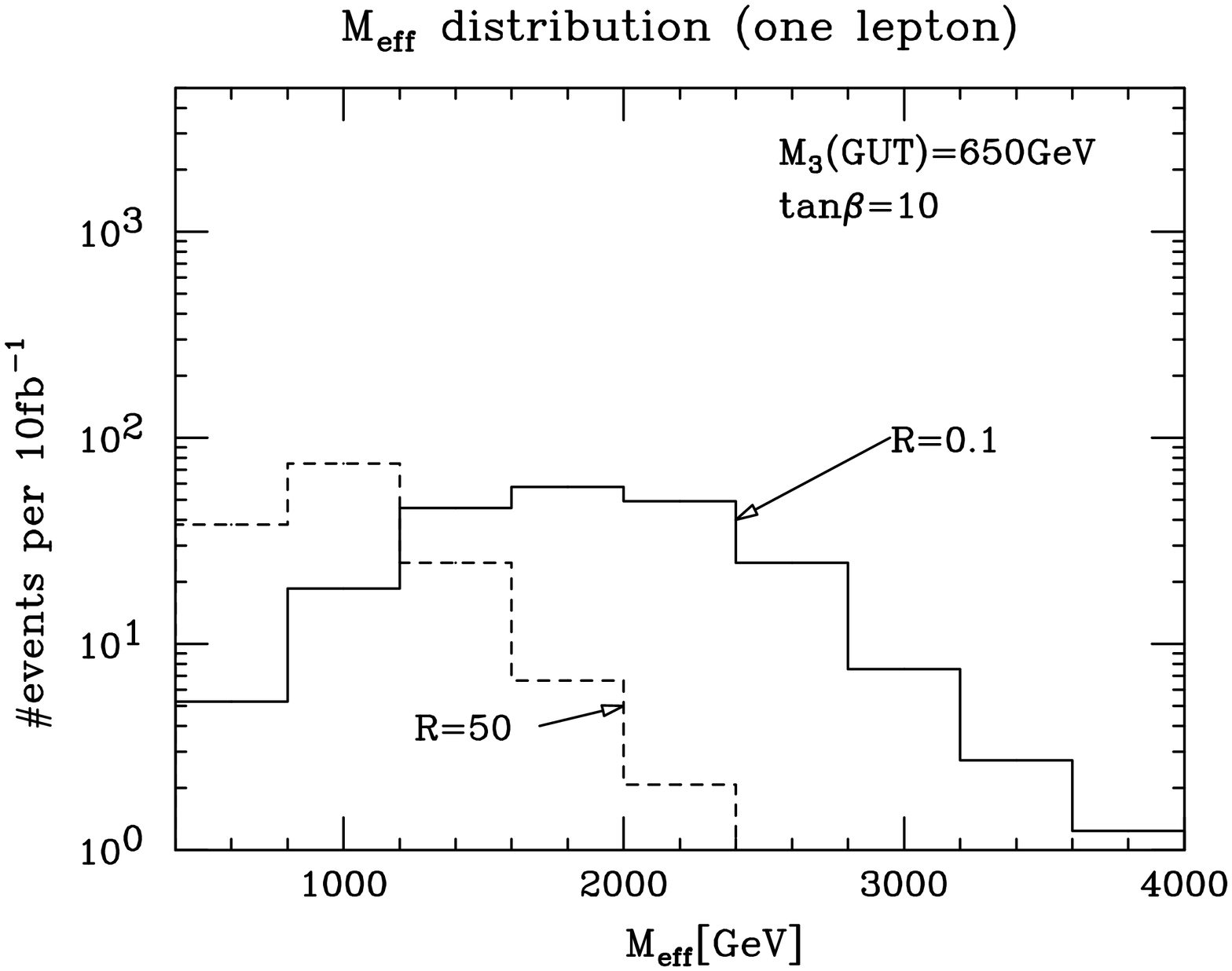}
\caption{The $M_{\rm eff}$ distributions for one 
 lepton signal events  for 
model B, $M_3{\rm (GUT)}=450$~GeV (left panel),   
$M_3{\rm (GUT)}=650$~GeV (right panel) 
and $\tan\beta=10$.       
Th  solid histograms for 
$R=0.1$ (MSUGRA like) and 
dashed histograms for $R=55$ and $50$ respectively 
corresponding to degenerated mass spectrum. 
All plots are normalized for  
 $\int L dt =10$fb$^{-1}$.  }
\label{fig:mefflB}
\end{figure}

\begin{figure}
\begin{center}
\includegraphics[width=6cm,angle=90]{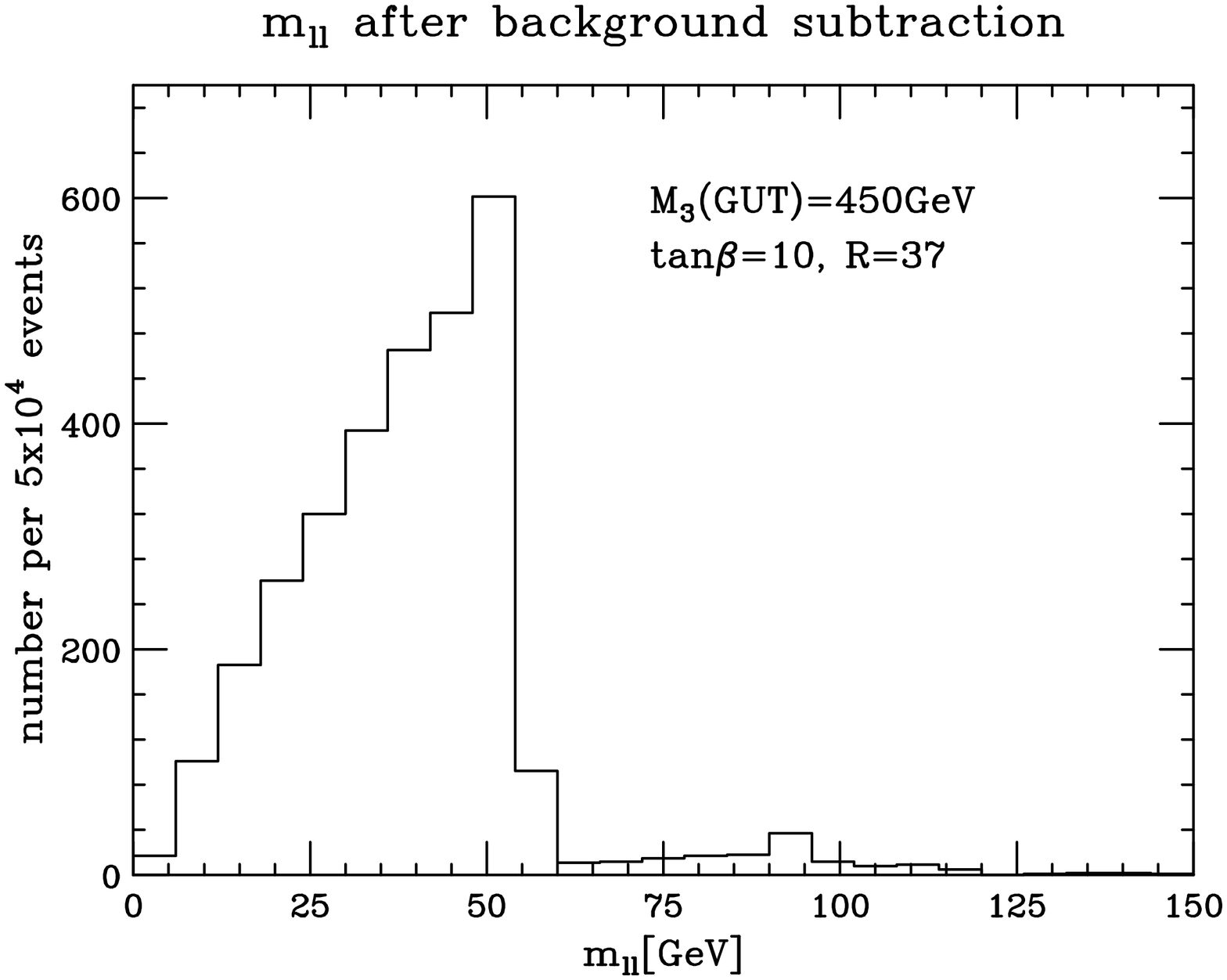}
\includegraphics[width=6cm, angle=90]{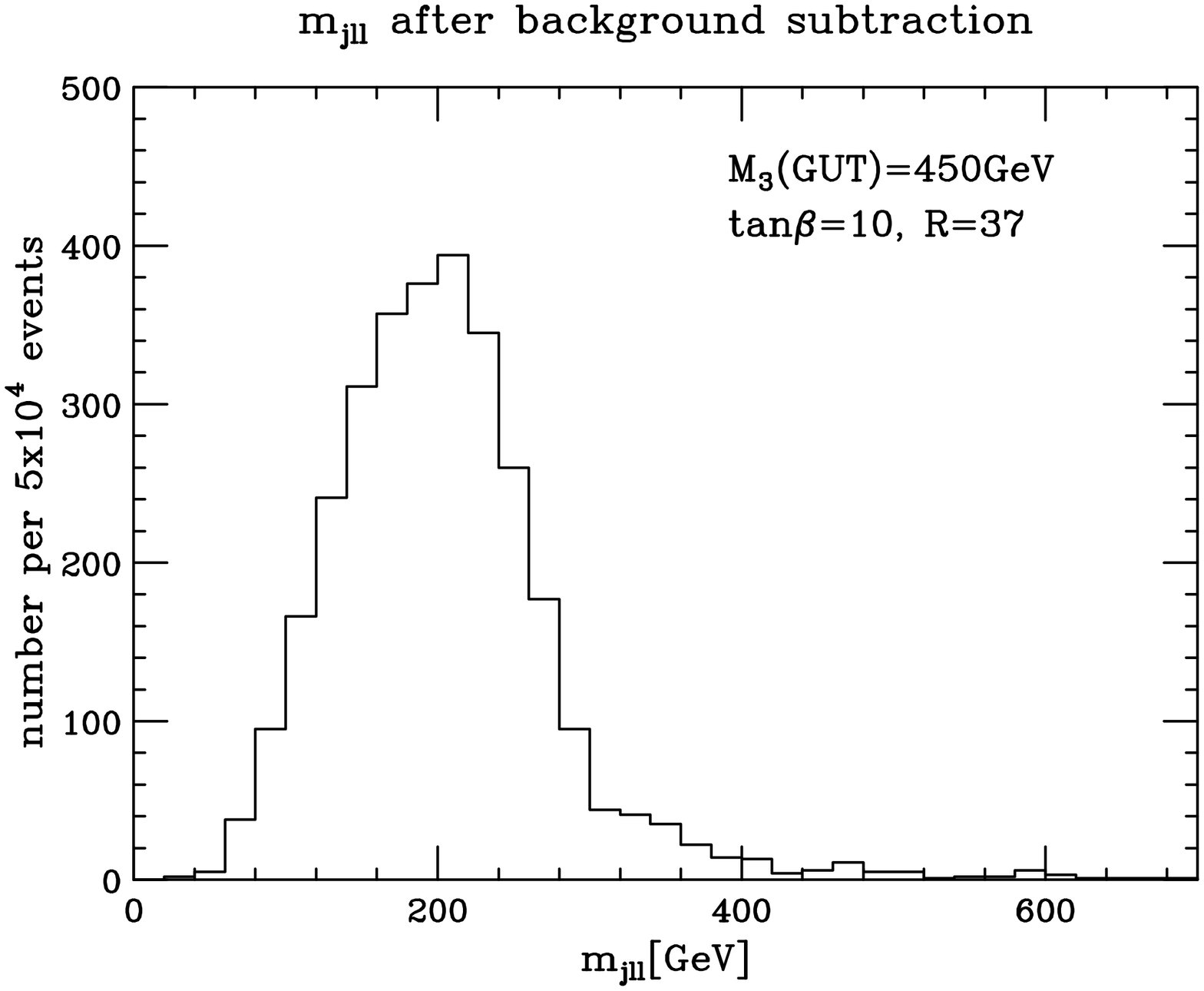}
\caption{The  $m_{ll}$ and $m_{jll}$ distribution of 
signal for a degenerated SUSY point.
$5\times 10^4$ SUSY events are generated for model A, 
$M_3$(GUT)$=450$~GeV, $\tan\beta=10$ and  $R=37$. }
\label{fig:mllA}
\end{center}
\end{figure}

The histograms in Figure~\ref{fig:mefflA} show signal 
distributions for 
$M_3{\rm (GUT)}=650$~GeV, $750$~GeV, $850$~GeV from 
left to right,  and $R\sim 0$ for solid line and $R=38, 37, 37$  
for dashed lines, respectively. 
As can be seen in Table~\ref{table:massAB}, 
the minimum mass difference is 
smaller  for the model B than that for the model A, 
because the higgsino mass 
$\mu$ is higher.  Because of the degeneracy, the signal 
is more suppressed as can be seen in Figure \ref{fig:mefflB}
where we take 
$M_3{\rm (GUT)}=450$~GeV (left) and $650$~GeV (right), 
$R=55$ and $50$ (dashed histograms), respectively.   
Overall normalization of 
the signal above $M_{\rm eff}> 1.6$~TeV reduces by a factor of 1/10  for  for model B
$M_3{\rm (GUT)}=650$~GeV (Figure \ref{fig:mefflB}, right panel)  
 compared to the model A  (Figure \ref{fig:mefflA}, left panel) 
 for most degenerate points.
We conclude that 
one lepton signal is not accessible at the most degenerated 
points by looking into $M_{\rm  eff}$ distribution 
alone, assuming that SM background in the one lepton channel 
in \cite{TEV4LHC}, and a factor 2$\sim$ 3 uncertainty, 
even though the SUSY scale is relatively small for this point, 
$m_{\tilde{g}}=1.48$~TeV  The mass 
reach at the LHC will be revisited again in section 4.5.

\subsection{Two lepton channel}
For the  model A, 
there are significant two  lepton events in  the $\mu\sim M_1$ region 
because $\tilde{l} $ is lighter than the $\widetilde{W}$-like neutralino. 
We show the $m_{ll} $  and $m_{jll}$ distribution 
for the model A,  with $M_{3}{\rm (GUT)}=450$~GeV and $R=37$
in Figure~\ref{fig:mllA}. 
Here we require 
that one of the two isolated leptons has $p_T>20$~GeV and the other 
has $p_T>10$~GeV. We also apply 
the standard cuts described in section~I. 
The relevant mass spectrum for the most prominent endpoints 
are $m_{\tilde{u}_R}=901$~GeV,
$m_{\tilde{d}_R}=869$~GeV,
$m_{\tilde{\chi}^0_2}=554$~GeV,
$m_{\tilde{\chi}^0_1}=499$~GeV,  and 
$m_{\tilde{e}_R}=524$~GeV. 
The predicted $m_{ll}$ edge and $m_{jll}$ endpoint 
for the decay chain 
$\tilde{q}_R\rightarrow \tilde{\chi}^0_2 \rightarrow 
\tilde{l}_R\rightarrow \tilde{\chi}^0_1$  is $m_{ll}=54.9$~GeV and
$m_{jll}=308$~GeV, respectively, consistent with Figure~\ref{fig:mllA}. 
The edges and endpoints of other cascade decays are also 
visible in the plots.  

To select the jet from $\tilde{q}$ decays, we have taken the one of 
two most highest $p_T$ jets with smaller $m_{jll}$. If mass difference 
between $m_{\rm \tilde{q}}$ and $m_{\tilde{\chi}^0_1}$ is smaller, 
we may found the smaller probability to find correct jets that 
arise from $m_{\tilde{q}}$ decay. We do not observe such 
effect in Figure~\ref{fig:mllA}. 

The number of accepted events $N_i$ 
(where  $i$ denotes  
number of isolated leptons with $p_T>20$~GeV
and $\vert \eta \vert <2.5$) for this point 
are $N_0=11229$, $N_1=5167$, 
$N_2=2645$ for $5\times 10^4$ generated events
(corresponding to $\sim$ 20~fb$^{-1}$) . 
Number of two lepton events below the dominant edge is 
2933  after the background subtraction.
Because $M_{\rm eff}$ distribution of  
the one lepton channel is not very prominent over the background,  
SUSY particles may be discovered in 
the two lepton channel for the degenerate region. 

Note that for the most degenerated SUSY spectrum, 
we have $M_1\sim M_2\sim \mu$ 
so that all neutralinos are highly mixed. 
In such situation,  dominant source of $\tilde{\chi}^0_2$
is $\tilde{q}_R$,  because 
$\tilde{q}_L$ has significant decay branching ratios 
into  charginos. It is worth to point out that 
the relative weight of the decays 
 $\tilde{q}_{L(R)}
\rightarrow \tilde{\chi}^0_2 \to \tilde{l}_{L(R)}$ may be 
studied by looking into the charge asymmetry in  
$m_{jl}$ distributions, $A\equiv (m_{jl^+}-m_{jl^-})$
$/(m_{jl^+}+m_{jl^-})$
\cite{BARR,GOTO}.  The charge asymmetry comes from the 
polarization of $\tilde{\chi}^0_2$ arising from $\tilde{q}$, 
however as discussed in \cite{WEBBER}, the effect tends to 
be suppressed if sparticle masses are degenerated. 

For the model B, the decay of $\tilde{\chi}^0_i$ 
into sleptons 
are closed for any value of $R$.   
For 
$M_3{\rm (GUT)}=450$~GeV and $R=40$, 
we have $N_0=13433$, $N_1=4412$, $N_2=642$   for 
$5\times 10^4$ generated events.
While $N_0$ and $N_1$ is about same as that of 
the model A, $N_2$ is reduced by a factor of 4.  In addition 
no significant events with 
opposite sign  same flavor leptons are seen 
after the subtraction of $e^{\pm} \mu^{\mp}$ events. 

\subsection{Discovery potential in one lepton channel}

In MSUGRA, the signal $M_{\rm eff}$ distribution 
is clearly harder than  the background. Especially, one 
can see the increase of the signal towards its peak, 
where the background is negligible.  If the 
total $M_{\rm eff}$ distribution shows a bump
or  a clear change in  the power behavior, one may 
claim discovery of SUSY particles without 
precise understanding of the background distribution. 
 
On the other hand, when SUSY mass spectrum 
is degenerated, the peak of the signal distribution 
shifts to a lower position where the background may 
be $\sim$10 times  higher. 
The signal distribution  shows
a similar power low behavior for 
$M_{\rm eff}> M^{\rm peak}_{\rm eff}$ 
as can be seen in Figure~\ref{fig:meffA}. The bump structure 
may not be detected 
and precise  understanding of the background 
would be  required.   Searching edges 
in $m_{ll}$ distribution from the decay $\tilde{\chi}^0_2\rightarrow$
$\tilde{l}\rightarrow \tilde{\chi}^0_1$ may become more important, 
however, as we have discussed already, the 
signal rate is  model dependent.  We do not consider the possibility to 
improve S/N ratio by selecting signal region $M_{eff}$ vs $\ETmiss$ 
in this subsection. 

Table~\ref{table:meffnum} shows  
the number of signal events after the cut 
for $M_{\rm eff}$ intervals for $5\times 10^4$ 
generated event (upper row) and number of events for 10~fb$^{-1}$
(lower row)  
for both model A and model B. 
We take a moderate value of $M_3{\rm (GUT)}=650$~GeV for the Table, 
which corresponds to $m_{\tilde{g}}\sim1.4$~TeV.

\begin{table}
\begin{tabular}{|ccccccc|c|c|}
\hline
$M_{\rm eff}$(TeV)&0.8-1.2&1.2-1.6&1.6-2.0&2.0-2.4&2.4-3.2
&3.2-4.0& $\sigma$&$2p_{\rm CM}$(TeV)\cr  
\hline
\hline
model A &&&&&&&&\cr
\hline
$R=0.1$&448&905&1267&999&568 &69&0.252pb&1.35\cr
 &  22.5& 45.6&63.9&50.3& 28.8& 3.4& &\cr
 \hline
10&752&1201&1378&896& 483&59&0.245pb&1.29\cr
& 36.8& 58.8 & 67.4 & 43.8& 23.9 & 2.9 & &\cr
\hline
20& 813& 1421& 1319& 676& 293&17& 0.248pb&1.20\cr
&40.3&70.5& 65.4&33.7& 14.5& 0.8& &\cr
\hline
30&  1246&1691&1018&336& 95& 9 &0.267pb& 1.04\cr 
&66.6&90.4& 54.4& 18.0&5.1& 0.5 &&\cr 
\hline
40& 2446&2337&758&212&63&9 & 0.314pb&0.87\cr 
& 152.3 & 145.5& 47.2& 13.2& 4.0& 0.6& &\cr 
\hline
50& 882& 960& 529& 198& 65& 4& 0.457pb& 0.90\cr
& 80.7&87.8& 48.4& 18.1& 5.9&0.4&& \cr 
\hline
55& 173& 249& 175& 71&26& 5 &0.666pb& 0.94\cr 
& 23.0& 33.2& 23.2 & 9.5& 3.5& 0.7& &\cr
\hline 
\hline
model B &&&&&&&&\cr
\hline
$R=0.1$ & 450& 1108& 1405&1192& 784& 96& 0.206pb&1.42\cr 
& 18.6& 45.7& 57.9& 49.2& 32.3& 4.0&& \cr
\hline
10&649& 1267& 1585& 1145& 654& 67&0.187pb& 1.38\cr
& 24.3& 47.4& 59.3&42.8& 24.5& 2.5&& \cr 
\hline
20& 836& 1528& 1645& 962& 432& 42& 0.174pb&1.31\cr 
& 29.1& 53.2& 57.2& 33.5& 15.0& 1.5& &\cr
\hline
30& 943&1838&1362&561& 217& 26&0.165pb&1.2\cr 
& 31.2& 60.7& 45.0 & 18.5& 7.2&0.9& &\cr
\hline
40& 1711& 1591& 711& 184& 86& 4& 0.158pb& 0.99\cr
&54.0& 50.2& 22.4& 5.8& 2.7& 0.1 && \cr
\hline  
50& 2428& 803& 215& 67& 25& 5& 0.154pb& 0.73 \cr
& 74.9& 24.8& 6.6& 2.1& 0.8 & 0.2& &\cr
\hline 
\end{tabular}
\caption{The number of events per $M_{\rm eff}$ bins in model A and 
model B. The upper rows show the events for 50000 events in our simulations, 
and lower rows show the event per 10~fb$^{-1}$. $M_3{\rm (GUT)}=
650$~GeV. }
\label{table:meffnum}
\end{table}

In the MSUGRA limit $(R \sim 0)$, the background rate is 
negligible in the signal region. 
The number of events in $M_{\rm eff}>2.5$~TeV 
is greater than 10 for $R<20$, where the number of 
background events would be around 1 according to 
\cite{TEV4LHC}. However, 
the signal rate is reduced by more than a factor of $4$  for
$R>30$ compared to that at $R\sim 0$.  
Another important question is if we have a flatter  power 
spectrum  in a certain $M_{\rm eff}$ region. 
The number of  background events is  reduced steeply 
$M_{\rm eff}$  bins in Figure \ref{fig:meffA}, while 
we have  nearly flat signal distribution between $1.2$-$2.4$~TeV 
for MSUGRA-like ($R\sim 0$)  points. 
The ratio 
between the signal in the bins 
$N(1.6{\rm TeV}<M_{\rm eff}<2.0{\rm TeV})$  
and $N(2.0{\rm TeV}<M_{\rm eff}<2.4{\rm TeV})  \sim3: 1$ 
for   $R=30, 40$ for the model A and for $R=30,40, 50$ for 
the model B, therefore 
the power law for the signal is similar to that of background 
in Figure \ref{fig:meffA} .
We can describe the discovery potential in terms of the 
mass degeneracy rather than $R$. In the Table~\ref{table:meffnum}, 
$2p_{\rm CM}$ ranges 
from $1.4\sim 0.7$~TeV.  ($M_{\tilde{q}}=1.4$~TeV ). 
Once $2p_{\rm CM}$ reduces below 1 TeV, 70\% of 
$M_{\tilde{q}}$, the 
separation between signal and background are not 
good in $M_{\rm eff}$ distributions.  
Good understanding of the 
background distribution and the detector effect is 
required to exclude the model in early stage of LHC runs
$\int dt {\cal L}=10$fb$^{-1}$.

A factor 2  decrease of the 
signal is found   for the model A 
between $R=50$ and $R=55$.   In this model,  squark mass 
decreases as $R$ increases. While the production cross section 
increases as 
${\tilde{q}}$ 
gets  lighter, the masses of 
heavier neutralino and chargino increase.  They are
heavier than $\tilde{q}$ at $R=55$, so that they would not 
be produced from $\tilde{q}$ decay at all. The lighter 
neutralinos and charginos 
are still lighter than $\tilde{q}$ but they are degenerated, 
so that only soft leptons are produced from the 
decay. 
The two lepton 
signal is also suppressed for this model point. This is an example 
that the rate of  signal events  with  lepton is  model dependent. 

The tendency of the signal reduction is same for the higher 
mass spectrum. The luminosity needed to discover the 
SUSY signal is larger at the degenerated SUSY points.  
For example, the production cross section is 
 $43.6$~fb for the model A with $M_3{\rm (GUT)}=850$~GeV  and $R=0$,
and  15.1 events/10~fb$^{-1}$ is expected in 
the region $2.4$~TeV $<M_{\rm eff}<4$~TeV, which may be enough to 
 claim  the discovery. On the other hand,  for $R=37$
only  3.8 events/10 fb$^{-1}$ is expected, excluding the possibility 
of discovery with low luminosity.

\section{discussion  and conclusion} 

In this paper, we study the discovery potential of 
minimal
supersymmetric models with degenerated  mass spectrum
at the LHC.
Such parameter regions have not been systematically
studied in the past. SUSY studies
have been mainly performed in the models where gluino and squark
are significantly heavier than the LSP, i.e.
in the MSUGRA model, gauge mediation model,
and so on.

We do not study 
the general MSSM model here, but take  mixed
modulas anomaly mediation (MMAM)  models, which
are parametrized by $R\propto F_C/F_T$, $\tan\beta$, and $M_3{\rm (GUT)}$.
The parameter $R$ is the ratio
of the $F$ terms  of the volume modulas $T$  in KKLT model and
the compensator field $C$ in $N=1$ MSUGRA.
This parameter $R$ gives us one parameter description from
the MSUGRA-like mass spectrum, where $M_3\gg M_1$,
to (moderately) degenerated mass spectrum.
The models  we
have studied include the MSSM points where
$m_{LSP}/M_{\tilde{q}, \tilde{g}}\sim 0.7$.

When sparticle mass spectrum is  degenerated,
energy of a particle from squark and gluino decays
is small relative to the mass scale. This is
reflected in the distributions of the
first $p_T$  jet, $\ETmiss$  and $M_{\rm eff}$.  They
tend to peak at smaller values  than MSUGRA
prediction for the same gluino and squark masses.
The background from the SM processes increases
rapidly for such low $p_T$ and low $M_{\rm eff}$
region, therefore, the SUSY signal has to
compete with large background. Indeed,
the signal
and background $M_{\rm eff}$ distributions are quite similar for
degenerated parameters we have studied.
We find the $S/N$ ratio can be far below 1
for our most degenerated parameter points. Nevertheless, 
we found the signal distribution of degenerated SUSY mass spectrum 
show special universal pattern in $M_{\rm eff}$ and $\ETmiss$ 
plane.  This  may help to determine the appropriate signal 
region and discriminate signal from background better. 

The discovery potential at the degenerated points therefore
depends on our knowledge on the background distribution.
The main background source to the SUSY productions is
multijet final states involving $W$, $ Z$ and $t$. There are
significant improvements in the lowest order estimation of the
background recently;  Matrix element 
corrections of the backgrounds are now included. 
Theoretical uncertainty comes
from matching between parton shower approximation
and matrix corrections, and  higher order QCD corrections (scale 
uncertainty). 

Experimentally  backgrounds are partly estimated  by the 
experimental data itself.  For example,  overall normalization of dominant 
$t\bar{t}$ background can be determined by looking into the 
events with low $\ETmiss$, where contamination of the 
SUSY events are small. 
For the degenerated mass spectrum, however, such calibration 
may also be affected.  Note that the background estimation 
in the SUSY signal region require extrapolation using Monte 
Carlo, while there is a contamination in the region for the 
calibration in non-trivial manner for degenerated SUSY mass spectrum. 
We look into $\ETmiss$ distribution 
and find that the number of events with $\ETmiss < 200$~GeV 
is increased by factor of $2$ for our most degenerated point 
 (the model B with $R=55$, $M_3{\rm (GUT)}=450$~GeV) from  
 MSUGRA limit ($R=0.1$). 

Even with the uncertainty discussed above,
the discovery of supersymmetry in MSUGRA models
would not be a  problem up to $\sim$
2~TeV region. However, as we have
discussed in this paper, this is not the case even
for  moderately degenerated SUSY parameters
in mixed modulas anomaly mediation model.
We stress the importance to include the
background and its uncertainty to perform
SUSY discovery and parameter studies
for  degenerated  or  general MSSM mass spectrum,
while we find that they tend to be ignored in recent studies
\cite{TATA, KANE}.

\section*{Acknowledgement}

We would like to thank  Dr. Tasuo Kobayashi, Dr.  Shoji  Asai and  Dr.  Junichi
 Kanzaki for discussion.  We also thank Dr.  Giacomo Polesello for comments 
 to the drafts. 
 This work is supported in part by Grant in Grant-in-Aid for Science Research, Ministry of Education, Science and Culture, Japan, 16081207, 18340060  for MMN and 
 15340076, 16081208 for KK.

\end{document}